\documentclass[12pt, a4paper]{article}
\pdfoutput=1
\usepackage{amsmath}
\usepackage{amsfonts}
\usepackage{amssymb}
\usepackage{graphicx, rotating}
\usepackage{epsfig}
\usepackage{latexsym}
\usepackage{graphicx}
\usepackage{color}
\usepackage{amsmath,bm,amssymb}
\usepackage{cite}
\usepackage{slashed}
\usepackage{hyperref}
\hypersetup{colorlinks, citecolor=bluscuro, linkcolor=black, urlcolor=bluscuro}
\definecolor{rossos}{cmyk}{0,1,1,0.55}
\definecolor{bluscuro}{rgb}{0.15, 0.2, .85}
\definecolor{bluchiaro}{cmyk}{1,.3,0.,0.1}

\numberwithin{equation}{section}

\newcommand{\be}{\begin{equation}}
\newcommand{\ee}{\end{equation}}
\newcommand{\bea}{\begin{eqnarray}}
\newcommand{\eea}{\end{eqnarray}}

\newcommand{\lsim}{\,\raisebox{-.1ex}{$_{\textstyle <}\atop^{\textstyle\sim}$}\,}

\newcommand{\arXiv}[2]{\href{http://arxiv.org/pdf/#1}{{\tt [#2/#1]}}}
\newcommand{\arXivold}[1]{\href{http://arxiv.org/pdf/#1}{{\tt [#1]}}}

\def\bma#1{\mbox{\boldmath{$#1$}}}

\begin{document}
\allowdisplaybreaks
%FRONTPAGE2%%%%%%
\begin{titlepage}
\begin{flushright}
IFT-UAM/CSIC-20-42
\end{flushright}
\vspace{.3in}

\vspace{1cm}
\begin{center}
{\Large\bf\color{black} 
Vacuum Decay in the Standard Model:\\
\vspace*{0.5cm}
 Analytical Results with Running and Gravity
} \\
\vspace{1cm}{
{\large J.R.~Espinosa}
%\vspace{0.3cm}
} \\[7mm]
{\it Instituto de F\'{\i}sica Te\'orica UAM/CSIC, \\ 
C/ Nicol\'as Cabrera 13-15, Campus de Cantoblanco, 28049, Madrid, Spain
}
\end{center}
\bigskip

\vspace{.4cm}

\begin{abstract} 
A tunneling bounce driving the decay of a metastable vacuum must respect an integral constraint dictated by simple scaling arguments that is very useful to determine key properties of the bounce. After illustrating 
how this works in a simple toy model, the Standard Model Higgs potential is considered, including quartic coupling running and
gravitational corrections as sources of scale invariance breaking. 
This approach clarifies the existence of the bounce and leads to
simple and accurate analytical results in an expansion in the breaking parameters.
 Using the so-called tunneling-potential approach (generalized for nonminimal coupling to gravity) the integral constraint and the tunneling action are extended to second order in perturbations.

\end{abstract}
\bigskip

\end{titlepage}

\section{Introduction\label{sec:introduction}}

The idea that a heavy chiral fermion could radiatively induce an instability in the potential of a light scalar is an old one \cite{SMold} that preceded both the discovery of the heavy top quark and of the light Higgs boson of the Standard Model (SM). Today, combining the precisely measured values of the top mass, $M_t=173.34\pm 0.76$ GeV \cite{mt}, and the Higgs boson mass, $M_h=125.09\pm 0.24$ GeV \cite{mh}
 with theoretical developments (three-loop renormalization group running, two-loop effective potential and matching) that allow an increasingly accurate extrapolation of the SM to higher energies it is likely that our vacuum is metastable \cite{SMnew} and  therefore is not the true vacuum of the theory.

Such metastable vacuum can decay by quantum tunneling via the nucleation of a bubble that probes the high-field region where the potential is deeper. This bubble subsequently grows at the speed of light transforming our low scale vacuum into the higher true one (actually crunching it \cite{CdL}). Nevertheless, the nucleation rate per unit volume, $\Gamma/V$, of such deadly bubbles is extremely small being exponentially suppressed as
\be
\Gamma/V = A e^{-S_E/\hbar}\ .
\ee
Here $S_E$ is the action of the Euclidean bounce that dominates the
decay \cite{Coleman}. As shown in detail later on, approximating the Higgs potential at high scales as $V(\phi)=-\lambda\phi^4/4$ 
the bounce can be obtained analytically (it is the so-called Fubini bounce) and $S_E=8\pi^2/(3\lambda)$ (we set $\hbar=1$ in the following). The prefactor $A$ includes the effects of fluctuations around the bounce \cite{ColemanCallan} and is much harder to calculate but has a subleading influence on the calculation of the rate. 

The tiny decay rate above should be multiplied by the huge size of the 4 dimensional volume of our past light-cone to arrive at the decay probability of our vacuum\footnote{Comparing this with proton decay experiments, we could say that our universe is performing a vacuum decay experiment that uses the largest container imaginable and has been running during the whole age of the Universe. Luckily the result of the experiment has been null so far.}, which turns out to be extremely small. In other words, the lifetime $\tau$ of our metastable vacuum is much larger than the age of the universe \cite{SMnew,IRS}. The most accurate estimates of this lifetime, including analytical calculations of the prefactor $A$ of the rate give $\tau> 10^{65}$ years at 95\% C.L. \cite{Matthew,CMS}.

Significant effort was needed to calculate the prefactor $A$, that plays a subleading role compared to the exponent in the decay rate
\cite{Matthew,CMS}. However, these works left out of the rate calculation gravitational corrections \cite{CdL} that can be as important as  the effect of the prefactor, given the fact that the characteristic scale of the bounce is not too far below the Planck scale. Gravitational corrections for the SM vacuum decay have been considered in several other works \cite{IRST,EFT,BMZ,RS,SSTU}. The calculation rests on a double assumption: first, that the Euclidean version of the gravitational action correctly describes vacuum decay and second,
 that the Euclidean bounce that extremizes the action has $O(4)$ symmetry (both assumptions can be proven in the absence of gravity).
Once this is assumed it is simple to solve numerically a system of two coupled differential equations for the Euclidean bounce and a single function describing the metric. The present paper is concerned instead with analytic approximations that give insight on the parametric dependence of these gravitational corrections to different quantities. 
This analytic attack was pioneered in \cite{IRST}, that treated gravitational corrections as perturbations over the Fubini solution and flat metric. This analytic approach was questioned in \cite{BMZ,RS} based on the (correct) observation that a potential with a negative quartic  potential plus gravity does not posses a bounce solution. Finally, \cite{SSTU} argued that the 
analytic approach of \cite{IRST} is justified on the basis that in the  SM potential the Higgs quartic coupling is not constant but a running coupling allowing the bounce to exist,  and obtained the analytic dependence of the gravitational corrections in the presence of a nonminimal coupling $\xi$ between the Higgs and the Ricci scalar.

The aim of this paper is to revisit the analytical calculation of gravitational corrections to the vacuum decay in the SM going beyond 
\cite{IRST,SSTU} in several respects. The existence or non existence of the bounce for different approximations to the SM potential is most easily and clearly understood by using a scaling argument used by Affleck in \cite{Affleck} to arrive at an integral constraint that the bounce should satisfy. This is important because having full analytic control over the conditions to have a proper bounce is more reassuring than having only a numerical solution of the problem. 
The method is of particular relevance for nearly scale invariant potentials with small perturbations that break exact scale invariance (precisely the case of the SM). 
Once the right ingredients to have a bounce are in place one can use a perturbative expansion in the breaking parameters to obtain explicitly the bounce and the action to the order desired. We illustrate all this in Section~\ref{sec:VmL} for a simple toy potential, used as a warm-up exercise. We also use the example to illustrate the general link between the Affleck constraint integral on the bounce and the extremality of the tunneling action.
In Section~\ref{sec:Vrunl}
we apply the method to a quartic potential with running coupling finding explicitly the first order perturbations induced by the running both on the bounce (a perturbation of the Fubini bounce)  and the tunneling action. In Section~\ref{sec:Vrunlgrav} we repeat the exercise adding gravity. The model has now all the ingredients to be a good approximation to the SM and one can already compare with the numerical results for the SM finding good agreement. 

In the sections described so far the Euclidean formulation of the tunneling bounce due to Coleman \cite{Coleman,CdL} is used. Section~\ref{sec:Vt} turns instead to the alternative formulation based on the so-called tunneling potential approach recently introduced in \cite{E}. First the formulation is extended to include a nonminimal coupling to gravity and then some general results are given for the perturbative calculation of the tunneling action over a scale invariant case. As the new method leads to simpler expressions compared with the Euclidean approach, we use it in Section~\ref{sec:high} to calculate the corrections to the tunneling action at second order in scale breaking.  Section~\ref{sec:conc} contains a summary and conclusions.

\section{The potential  $\bma{V(\phi)=m^2\phi^2/2-\lambda\phi^4/4+\phi^6/\Lambda^2}$\label{sec:VmL}}

Consider the potential 
\be
V(\phi)=\frac12 m^2\phi^2-\frac{\lambda}{4}\phi^4+\frac{1}{\Lambda^2}\phi^6\ ,
\label{VmlL}
\ee
with $\Lambda^2\gg m^2>0$.
For $m^2/\Lambda^2<\lambda^2/32$
the origin at $\phi_+=0$ is a metastable vacuum separated by a barrier from the true vacuum located at $\phi_-$ with $\phi_-^2=\Lambda^2\left[\lambda+\sqrt{\lambda^2-24m^2/\Lambda^2}\right]/12$.

\subsection{General Considerations}
Assume there is a bounce $\phi_B(r)$ for the decay out of $\phi_+$ and consider the rescaled field profile $\phi_a(r)\equiv a \phi_B(ar)$. The Euclidean action for the rescaled field, after changing the integration variable, reads
\be
S_E[\phi_a]=2\pi^2\int_0^\infty \left[\frac12 \left(\frac{d\phi_B}{dr}\right)^2-\frac14 \lambda\phi_B^4\right]r^3dr +2\pi^2
\int_0^\infty \left(\frac{1}{2a^2} m^2\phi_B^2+\frac{a^2}{\Lambda^2}\phi_B^6\right)r^3dr \ .
\ee
As $\phi_B(r)$ is by assumption a bounce, it extremizes the 
Euclidean action and, therefore, one should have $dS_E[\phi_a]/da=0$ at $a=1$, which translates into the integral constraint
\be
\boxed{
\int_0^\infty \left(-\frac12 m^2\phi_B^2+\frac{1}{\Lambda^2}\phi_B^6\right)r^3dr=0}\ .
\label{rescaling}
\ee 
From this condition we see that both the mass term and the sixtic are necessary for the bounce to exist.\footnote{Nevertheless, vacuum decay can still proceed in the absence of a bounce, see \cite{En} for a recent discussion.} Indeed, using the undershoot-overshoot method \cite{Coleman} to find the bounce one finds that for $m^2\neq 0$ and $\Lambda\rightarrow\infty$ there are only undershots, while $m^2=0$ with a finite $\Lambda$ leads only to overshots. 
We can also estimate from (\ref{rescaling}) that $\phi_0\equiv \phi_B(0)$ should scale as
\be
\phi^4_0\sim m^2\Lambda^2\ .
\label{phi0est}
\ee

For the potential (\ref{VmlL}), dimensional analysis also tells us that the tunneling action, which is dimensionless, must be a function of the ratio $m^2/\Lambda^2$:
\be
S_E=S_E(\lambda, m^2/\Lambda^2)\ ,
\ee
and we can infer the limiting values of the function $S_E$ as follows.
When $m/\Lambda\rightarrow 0$ we should recover the action for the pure (negative) quartic potential, for which the bounce is calculable analytically and given by the Fubini instanton \cite{Fubini,Lipatov}
\be
\phi_F(r)=\frac{\phi_0}{1+\lambda\phi_0^2r^2/8}\ ,
\label{Fubini}
\ee
which gives
\be
S_E(\lambda,0)= \frac{8\pi^2}{3\lambda}\ .
\ee
On the other hand, for $m^2/\Lambda^2=\lambda^2/32$ the
broken minimum $\phi_-$ of the potential (\ref{VmlL}) is degenerate with the minimum at the origin so that no decay is possible and the tunneling action becomes infinite. Therefore
\be
S_E(\lambda,\lambda^2/32)=\infty\ .
\ee

In what follows we calculate the first term in an expansion of $S_E$ in powers of $m/\Lambda$
\be
S_E(\lambda,m^2/\Lambda^2) = S_0 + S_1 +... 
\label{SmLexp}
\ee 
with $S_0=8\pi^2/(3\lambda)$ and $S_n$ being ${\cal O}(m^n/\Lambda^n)$.

\subsection{Perturbative Analysis}
For nonzero values of $m^2$ and $\Lambda^2$ one should find the bounce $\phi_B(r)$ by solving the Euler-Lagrange equation for extremals of  the Euclidean action \cite{Coleman}
\be
\ddot \phi +\frac{3}{r}\dot\phi = V'\ ,
\label{bounceq}
\ee
where dots (primes) denote derivatives with respect to $r$ ($\phi$) and the boundary conditions are $\phi(0)=\phi_0$, $\dot\phi(0)=0$, $\phi(\infty)=\phi_+$, with $\phi_0$ to be determined.
We can find approximate solutions for the bounce, that deviates from the Fubini profile (\ref{Fubini}), considering the mass term and the sixtic term as perturbations that break scale invariance. We consider them to be of the same order, $\epsilon_0^2$, as indicated by (\ref{rescaling}). The analysis requires that we look at the bounce configuration in two separate field regimes.

Consider first the bounce at small values of $r$ (or large values of $\phi$), for which
\be
\frac12 m^2\phi^2 \ll \frac14 \lambda\phi^4 \gg \frac{1}{\Lambda^2}\phi^6\ .
\ee
The first inequality requires
\be
\phi\gg \sqrt{\frac{2}{\lambda}}m\ .
\ee
The second inequality is guaranteed to hold, as 
\be
\phi\leq \phi_0\sim \sqrt{m\Lambda}\ll \Lambda\ .
\ee
In this regime we can approximate the bounce as
\be
\phi_L(r)=\frac{\phi_0}{1+\lambda\phi_0^2r^2/8}+ \delta\phi_L(r)\ ,
\ee
where the subscript $L$ refers to the large field regime. The 
term $\delta\phi_L(r)$ measures the deviation of the bounce from the Fubini configuration at first order in $m^2$ and $1/\Lambda^2$ ($\sim \epsilon_0^2$). Notice also that $\phi_0$ is at this point an unknown to be fixed eventually in terms of $m^2$ and $\Lambda^2$.
By expanding the bounce equation (\ref{bounceq}) at first order in $m^2$ and $1/\Lambda^2$ and solving for $\delta\phi_L(r)$ with the boundary conditions $\delta\phi_L(0)=0$, $d\delta\phi_L/dr(0)=0$ one finds
\bea
\delta\phi_L&=&\frac{\phi_0}{\lambda(1+x)^2}\left\{
\frac{\phi_0^2}{5\Lambda^2}\left[\frac{x}{1+x}\left(
24+13x-x^2\right)+6(1-x)\ln(1+x)
\right]
\right.\\
&+&
\left.\frac{2m^2}{\phi_0^2}\left[-1+16x-3x^2+\left(\frac{1}{x}-9-9x+x^2\right)\ln(1+x)+6(1-x)\mathrm{Li}_2(-x)\right]\right\}\ ,\nonumber
\eea
with $x\equiv \lambda\phi_0^2r^2/8$.

In the small-field regime, with $\phi\ll \sqrt{2/\lambda}m$, the mass term dominates over the other terms of the potential and the Fubini bounce is no longer a good starting point for the perturbative expansion. Keeping also subleading terms from the quartic, needed for the correct matching, one finds that the bounce is approximated by
\be
\phi_S(r)=\frac{8m}{\lambda\phi_0 r}K_1(m r) -\frac{512m^3}{\lambda\phi_0^3r}\int_{m r}^{\infty}\frac{K_1(z)^3}{z}\left[K_1(m r)I_1(z)-K_1(z)I_1(m r)\right]dz\ ,
\ee
where the subscript $S$ refers to the small field regime. The functions
 $I_1(z)$ and $K_1(z)$ are the modified Bessel functions of the first and second kind, respectively.

\begin{figure}[t!]
\begin{center}
\includegraphics[width=0.5\textwidth]{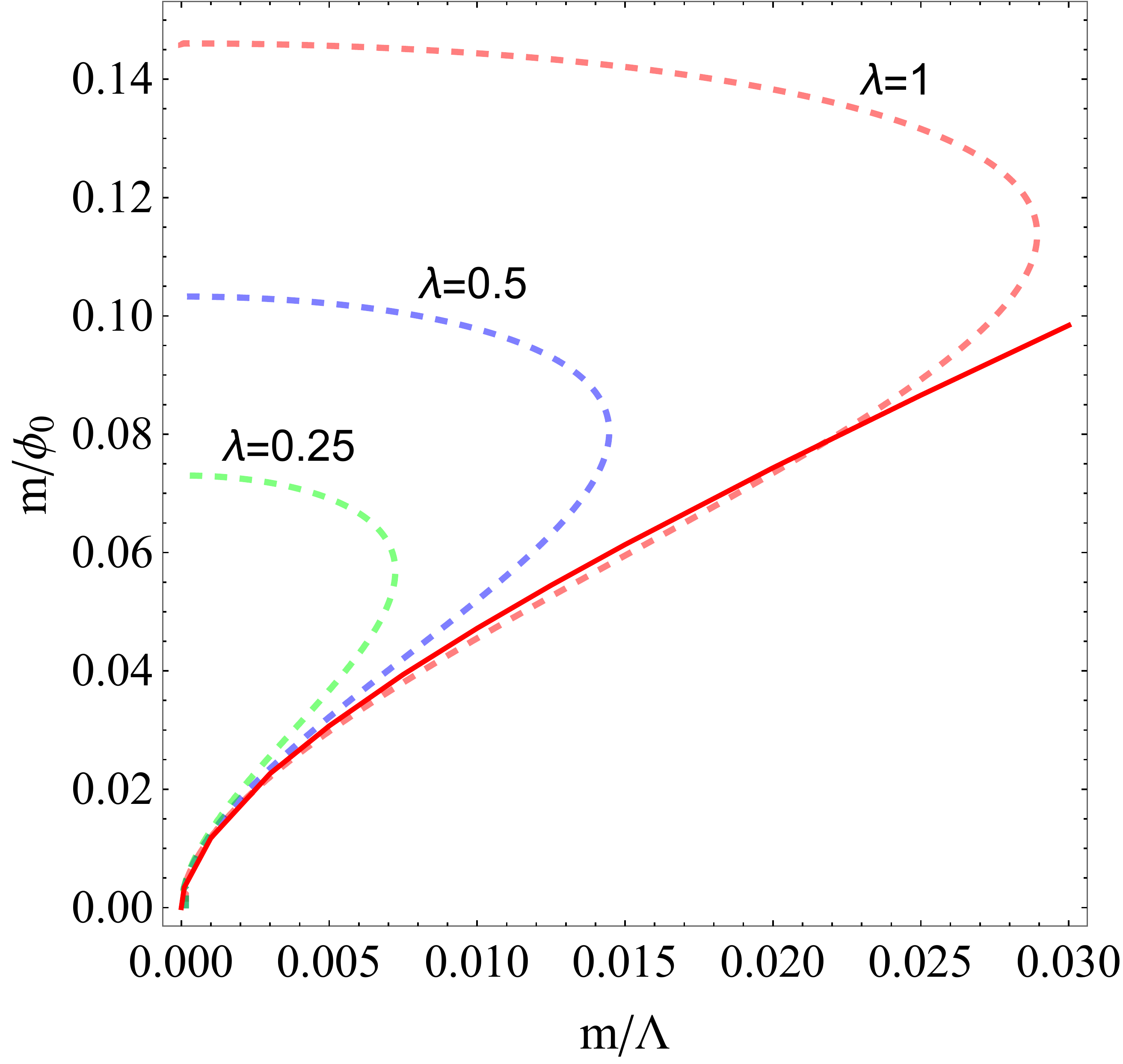}
\end{center}
\caption{Dashed lines: $m/\phi_0$ computed from Eq.~(\ref{phi0}) as a function of $m/\Lambda$ for the indicated values of $\lambda$. Solid line: 
$m/\phi_0$ from the numerical solution for the bounce, with $\lambda=1$.
\label{fig:phi0}
}
\end{figure}

Small and large-field regimes meet at $\phi(r_x)\simeq\sqrt{2/\lambda}\, 
 m$
for
\be
r_x^2=\frac{4}{\phi_0 m}\sqrt{\frac{2}{\lambda}}\ .
\ee
To match $\phi_L(r)$ and $\phi_S(r)$ at $r_x$ we use the 
large $r$ expansion of $\phi_L$
\be
\phi_L(r_x)\simeq  \frac{1}{\lambda\phi_0}\left[\frac{8}{r_x^2}-\frac{64}{\lambda_0\phi_0^2r_x^4}+2m^2\left(\ln\frac{\lambda\phi_0^2r_x^2}{8}-3\right)
-\frac{\phi_0^4}{5\Lambda^2}\right]+{\cal O}\left(\epsilon_0^3\right)\ ,
\ee
where we drop terms that contribute to the matching (at $r_x\sim 1/\sqrt{m}$) at order higher than $m^2$ and $1/\Lambda^2$ ($\sim\epsilon_0^2$). On the other hand, the small $r$ expansion of $\phi_S$ gives
\be
\phi_S(r_x)\simeq \frac{1}{\lambda\phi_0}\left[\frac{8}{r_x^2}-\frac{64}{\lambda_0\phi_0^2r_x^4}+2m^2\left(-1+2\gamma_E+\ln\frac{m^2r_x^2}{4}\right)
+{\cal O}(\epsilon_0^3)\right]\ ,
\ee
where $\gamma_E\simeq 0.577216$ is Euler's constant.
The matching $\phi_L(r_x)=\phi_S(r_x)$ gives an implicit formula for $\phi_0$ 
\be
\phi_0^4 \simeq 10\ m^2\Lambda^2\left(\ln\frac{\lambda\phi_0^2}{2m^2}-2-2\gamma_E\right) \ ,
\label{phi0}
\ee
that confirms the scaling anticipated in (\ref{phi0est}). It can be checked that the 
relation (\ref{rescaling}) is also satisfied for the obtained $\phi_L(r)$ and $\phi_S(r)$ with $\phi_0$ as given 
above. It is interesting in this example that the resulting value of $\phi_0$ comes from a UV/IR interplay of the operators $m^2\phi^2$
(that dominates in the IR) and $\phi^6/\Lambda^2$ (that dominates in the UV) and both are needed to produce the bounce.

Figure~\ref{fig:phi0} shows the solution of (\ref{phi0}) for different values of $\lambda$ as a function of $m/\Lambda$ (dashed lines). For fixed $m/\Lambda$ there are two solutions for $\phi_0$. We can chose the right branch by noting that for $\Lambda\rightarrow\infty$ with fixed $m$, we only have undershots, which implies $\phi_0\rightarrow\infty$. Therefore, we should choose the branch that has $m/\phi_0\rightarrow 0$ for $m/\Lambda\rightarrow 0$. This is confirmed by comparison with $\phi_0$ obtained via the numerical solution for the bounce, shown as a solid red line for the case $\lambda=1$. We also see from this comparison at which point the analytical approximation from (\ref{phi0}) would need to be corrected for larger $m/\Lambda$.

\begin{figure}[t!]
\begin{center}
\includegraphics[width=0.5\textwidth]{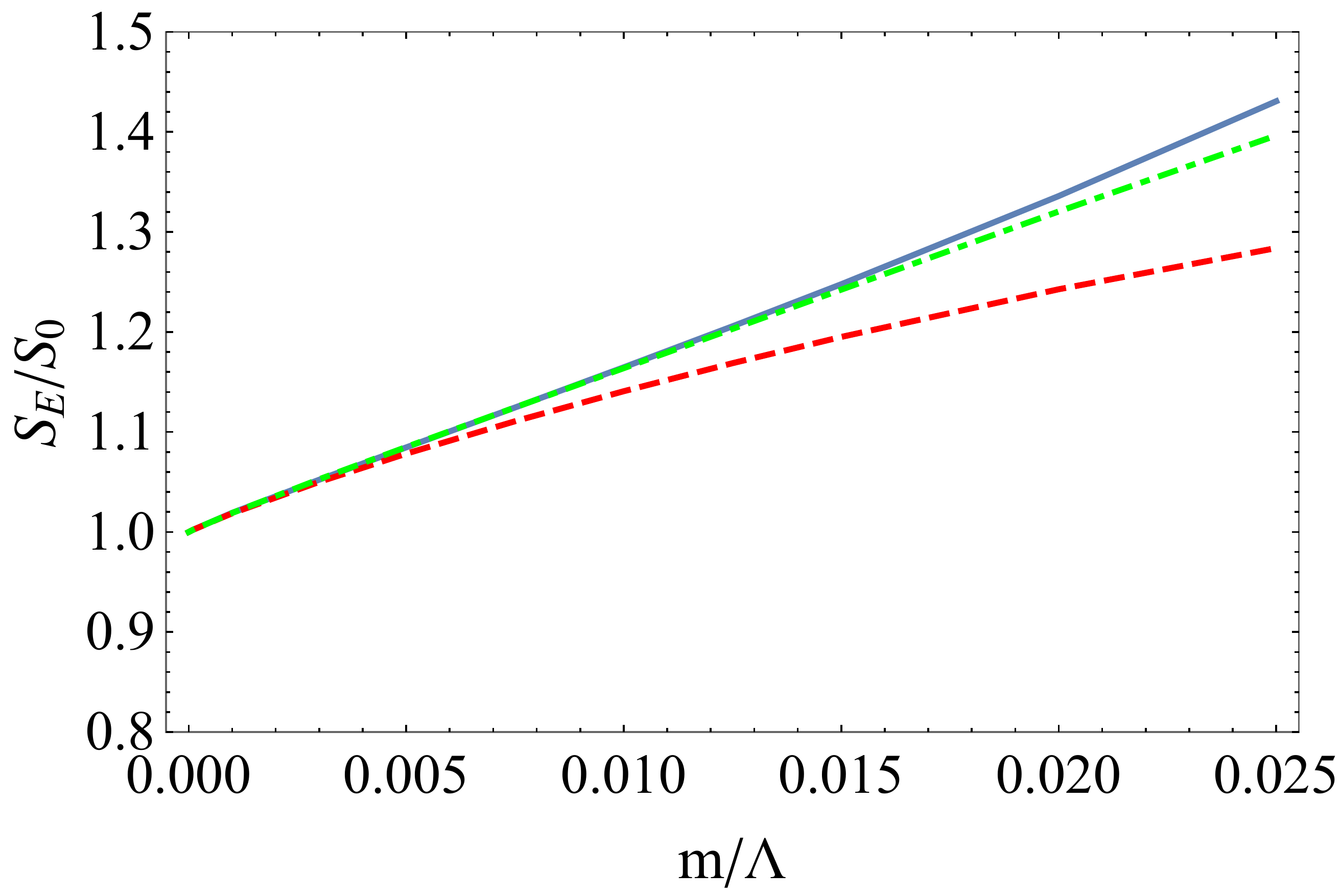}
\end{center}
\caption{Tunneling action [normalized to $S_0= 8\pi^2/(3\lambda)$], for the potential (\ref{VmlL}) with $\lambda=1$, calculated numerically (solid line) 
and approximated by Eq.~(\ref{SmLap}), $S_E/S_0\simeq (S_0+S_1)/S_0$ (dashed line). The dot-dashed line gives the resummed approximation $1/(1-S_1/S_0)$.
\label{fig:SmL}
}
\end{figure}

The tunneling action can then be  computed using the previous approximations for the bounce solution. To compute this action we split the $r$ integration interval into a low-$r$ subinterval $(0,a r_x)$, where we use $\phi_L(r)$ and a high-$r$ subinterval $(a r_x,\infty)$, where we use $\phi_S(r)$. Here $a$ is and ${\cal O}(1)$ arbitrary constant useful to check that the final result does not depend on the precise location of the matching point. One arrives at
\bea
S_E &\simeq &\frac{8\pi^2}{3\lambda}+\frac{32\pi^2 m^2}{\lambda^2\phi_0^2}\left(3+2\gamma_E-\ln\frac{\lambda\phi_0^2}{2m^2}\right)+\frac{48\pi^2\phi_0^2}{5\lambda^2\Lambda^2}+{\cal O}(m^2/\Lambda^2)
\label{SmLap0}\\
&=&
\frac{8\pi^2}{3\lambda}+\frac{32\pi^2}{\lambda^2}\left(\frac{m^2}{\phi_0^2}+\frac{\phi_0^2}{5\Lambda^2}\right)+{\cal O}
(m^2/\Lambda^2)\ ,
\label{SmLap}
\eea
where we have used (\ref{phi0}) to write the last expression. The terms kept are ${\cal O}(m/\Lambda)$ as $\phi_0^2\sim m \Lambda$, so we have computed $S_1$ in the expansion (\ref{SmLexp}).
Fig.~\ref{fig:SmL} shows a comparison of this approximation (dashed red line) with the numerical result (solid line), for $\lambda=1$, normalized to $S_0= 8\pi^2/(3\lambda)$, that is $S_E/S_0\simeq 1+S_1/S_0$. To illustrate the numerical impact of higher order corrections [of order ${\cal O} (m^2/\Lambda^2)$] we also plot (dot-dashed line) the resummed quantity $1/(1-S_1/S_0)$, which does a good job in approximating the numerical $S_E/S_0$ (although there is no theoretical justification for this).

\subsection{A General Result\label{subsec:link}}
An interesting property of the analytic action (\ref{SmLap0}) is that it is an extremal at the $\phi_0$ that satisfies the constraint relation (\ref{phi0}). In other words,
the condition $\partial S_E/\partial\phi_0=0$ gives eq.~(\ref{phi0}). Usually one expects this to happen for deformations of the bounce profile 
but this case is special because changing $\phi_0$ does not correspond to changing a trial profile $\phi(r)$. However, there is a deeper reason for this property to hold in general. If we consider the Euclidean action
evaluated on the bounce, its derivative with respect to $\ln\phi_0$ can be written as
\bea
\frac{\partial S_E[\phi_B]}{\partial \ln\phi_0}&=&\int_0^\infty
\left.\frac{\delta S_E}{\delta\phi}\right|_{\phi_B}\frac{\partial\phi_B}{\partial \ln\phi_0}\, dr\nonumber\\
&+&2\pi^2\int_0^\infty \left(\frac12\frac{\partial m^2}{\partial \ln\phi_0}\phi_B^2+\frac14\frac{\partial \lambda}{\partial \ln\phi_0}\phi_B^4-\frac{1}{\Lambda^4}\frac{\partial \Lambda^2}{\partial \ln\phi_0}\phi_B^6\right)r^3dr\ . 
\label{dSEdt0}
\eea
The first term of the right hand side vanishes due to the Euler-Lagrange equation satisfied by the bounce, $\delta S_E/\delta\phi=0$. The parameters of the potential have an implicit dependence on $\phi_0$ as this quantity is determined in terms of them, as in (\ref{phi0}). Simply, if all massive parameters in the potential are rescaled by some factor, $\phi_0$ would be rescaled by that same factor.
It follows that the derivatives of the potential parameters are simply given by the corresponding engineering dimensions
\be
\frac{\partial m^2}{\partial \ln\phi_0}=2m^2\ ,\quad
\frac{\partial \lambda}{\partial \ln\phi_0}=0\ , \quad
\frac{\partial \Lambda^2}{\partial \ln\phi_0}=2\Lambda^2\ .
\ee
Simple dimensional analysis guarantees that (\ref{phi0}) is consistent with these derivatives. Moreover, plugging them in (\ref{dSEdt0})
we get that $\partial S_E[\phi_B]/\partial \ln\phi_0$ is proportional to the constraint (\ref{rescaling}) and therefore vanishes.  In fact, the simplest way of obtaining the integral constraint on the bounce is to take derivatives of the action with respect to $\ln\phi_0$ as done above. It is also clear that the relation between $\partial S_E[\phi_B]/\partial \phi_0=0$ and the integral constraint on the bounce is a general result that holds beyond the particular example we have used to illustrate it and follows simply from dimensional analysis. One then has
\be
\boxed{
\frac{\partial S_E[\phi_B]}{\partial \phi_0}=0\quad \Rightarrow
\quad
\int_0^\infty \sum_\alpha\frac{\partial V(\phi_B)}{\partial p_\alpha}d_{(\alpha)} p_\alpha  r^3dr=0}\ ,
\label{link}
\ee
where $ p_\alpha$ are the parameters in the potential $V$, with engineering dimensions $d_{(\alpha)}$.

\section{The potential  $\bma{V(\phi)=\lambda(\phi)\phi^4/4}$\label{sec:Vrunl}}

Consider next the potential 
\be
V(\phi)=\frac{1}{4}\lambda(\phi)\phi^4\ ,
\label{Vrunl}
\ee
with a running quartic coupling.
For simplicity we take
\be
\lambda(\phi)= \lambda(\mu) + \beta_\lambda(\mu)\ln(\phi/\mu)+\frac12\beta_\lambda'(\mu)\ln^2(\phi/\mu)\ ,
\ee
where $\mu$ is some reference scale and $\beta'_\lambda=d\beta_\lambda/d\ln\mu$. As ultimately we are interested in the SM case, we consider values of $\lambda$, $\beta_\lambda$ and $\beta'_\lambda$ that imitate the behaviour of the running quartic in the SM. In particular we start with a positive
quartic at low energy $\lambda(\mu_{\mathrm IR})>0$ that becomes negative at some high scale due to $\beta_\lambda(\mu_{\mathrm IR})<0$ but eventually turns positive again due to 
$\beta'_\lambda>0$.
To simplify the analysis, make the logarithms smaller and have the Fubini instanton as a good zero-th order approximation to the bounce it is convenient to choose a renormalization scale close to the scale where $\lambda$ is negative.\footnote{We implicitly assume that we are not close to the critical Higgs mass point, for which the potential has nearly degenerate vacua, and the minimal value for $\lambda$ is very small. In that case the bounce is thin-walled and not close to a Fubini instanton.} In particular we use the scale $\mu_0$ at which $\beta_\lambda(\mu_0)=0$. At that scale 
$\lambda(\mu_0)\equiv -\lambda_0<0$. Omitting the implicit scale dependence we then write
\be
\lambda(\phi)= -\lambda_0+\frac12\beta_\lambda'\ln^2(\phi/\mu_0)\ .
\ee
As an example, for $M_h=125.09$ GeV and $M_t=173.34$ GeV [with $\alpha_S(M_Z)=0.1184$] one gets $\lambda_0=0.0143$, $\beta_\lambda'=5.6\times 10^{-5}$ and $\mu_0/m_P=0.59$, where $m_P=2.435\times 10^{18}$ GeV is the reduced Planck mass. 

\subsection{General Considerations}
As in the previous section, we can find approximate solutions for the bounce and the tunneling action considering the $\beta_\lambda'$ term as a 
perturbation that breaks scale invariance: due to $\beta_\lambda'\neq 0$ the bounce configuration deviates from the Fubini
bounce and the tunneling action deviates from 
$8\pi^2/(3\lambda_0)$. 

If $\phi_B(r)$ is the bounce, consider the rescaled field profile $\phi_a(r)\equiv a \phi_B(ar)$. The Euclidean action for the rescaled field configuration, after changing the integration variable, reads
\be
S_E[\phi_a]=2\pi^2\int_0^\infty \left[\frac12 \left(\frac{d\phi_B}{dr}\right)^2-\frac14 \lambda_0\phi_B^4\right]r^3dr +2\pi^2
\int_0^\infty \frac{1}{8} \beta_\lambda' \phi_B^4\left(
\ln\frac{a \phi_B}{\mu_0}\right)^2 r^3dr \ .
\ee
As $\phi_B(r)$ extremizes the Euclidean action we shoud have $dS_E[\phi_a]/da=0$ at $a=1$, which gives
\be
\boxed{
\beta_\lambda' \int_0^\infty  \phi_B^4 \ln(\phi_B/\mu_0)\, r^3dr=0}\ .
\label{rescalingrun}
\ee 
The same condition is obtained from $dS_E/d\phi_0=0$, using (\ref{link}) as explained at the end of the previous section, with
\be
\frac{\partial \lambda_0}{\partial \ln\phi_0}=0\ , \quad
\frac{\partial \beta_\lambda'}{\partial \ln\phi_0}=0\ , \quad
\frac{\partial \mu_0}{\partial \ln\phi_0}=\mu_0\ .
\label{dims}
\ee

In order to fulfill the condition (\ref{rescalingrun}) the integrand should change sign
and we learn that
\be
\phi_0\equiv\phi_B(0)>\mu_0\ .
\label{phi0exp}
\ee
As $\mu_0$ is the only mass scale in the problem, we have $\phi_0=c\mu_0$, and the above condition is $c>1$. At first order in $\beta_\lambda'$,
plugging in (\ref{rescalingrun}) the Fubini profile $\phi_{F}(r)$ for $\phi_B$
we get $\phi_0=\mu_0 e^{5/6}$. This result is confirmed below.

For this potential (\ref{Vrunl}), dimensional analysis also tells us that the tunneling action, which is dimensionless, must be a function
\be
S_E=S_E(\lambda_0,\beta_\lambda')\ ,
\ee
with no explicit dependence on the scale $\mu_0$.
When $\beta_\lambda'\rightarrow 0$ we should recover the action for the pure (negative) quartic potential, so
\be
S_E(\lambda_0,0)= \frac{8\pi^2}{3\lambda_0}\ .
\ee
On the other hand, the thin-wall limit of infinite action is only reached for $\lambda_0/\beta_\lambda'\rightarrow 0$. Therefore
$S_E(\lambda_0,\infty)=\infty$,
although we are not interested in this limit.

\subsection{Perturbative Analysis}
As was done in the previous section, we look at the bounce configuration in two separate field regimes. Consider first the bounce at small values of $r$, or large values of $\phi\sim \mu_0$, for which
\be
 \lambda_0 \gg \beta_\lambda'\ln\frac{\phi}{\mu_0}\ .
\ee
This is satisfied for $\phi\gg \phi_x$ with
\be
\phi_x\equiv \mu_0\ e^{-\sqrt{2\lambda_0/\beta_\lambda'}}\ll \mu_0 \ .
\ee
In this regime we approximate the bounce as
\be
\phi_L(r)=\frac{\phi_0}{1+\lambda\phi_0^2r^2/8}+ \delta\phi_L(r)\ ,
\ee
where the subscript $L$ refers to the large field regime. The 
term $\delta\phi_L(r)$ measures the deviation of the bounce from the Fubini configuration at first order in $\beta_\lambda'$. At this point $\phi_0$ is an unknown to be fixed eventually in terms of $\mu_0$.
By expanding the bounce equation at first order in $\beta_\lambda'$ and solving for $\delta\phi_L(r)$ with the boundary conditions $\delta\phi_L(0)=0$, $d\delta\phi_L/dr(0)=0$ we find
\bea
\delta\phi_L(r)&=&\frac{\beta_\lambda'\phi_0}{12\lambda_0(1+x)^2}\left[
(9-x)x\ln\frac{\phi_0}{\mu_0}+3(x-1)\ln^2\frac{\phi_0}{\mu_0}+\left(4+5x-\frac{1}{x}\right)\ln(1+x)
\right.\nonumber\\
&+&
\left.
3(1+x)\ln^2\frac{\phi_0}{(1+x)\mu_0}+1+\frac16(5x-63)+6(x-1)\mathrm{Li}_2(-x)\right]\ ,
\label{deltaphiL}
\eea
with $x\equiv \lambda_0\phi_0^2r^2/8$.

In the small-field regime, with $\phi\ll \phi_x$, linearizing the bounce equation we find that the bounce is approximated by
\be
\phi_S(r)=\frac{C_\phi}{r^2}\ ,
\ee
where the subscript $S$ refers to the small field regime.

Small and large-field regimes meet at $\phi(r_x)=\phi_x$
for
\be
r_x^2\simeq \frac{8}{\lambda_0\phi_0 \phi_x}\ .
\ee
To match $\phi_L(r)$ and $\phi_S(r)$ at $r_x$ we use the 
large $r$ expansion of $\phi_L$:
\be
\phi_L(r)\simeq  -\frac{\beta_\lambda'\phi_0}{12\lambda_0} \left(L_0-\frac56\right)+\frac{\phi_0}{x}+
\frac{\beta_\lambda'\phi_0}{2\lambda_0 x}\left[
\left(L_0+\frac{11}{6}\right)L_0- \left(L_0-\frac56\right)\ln x-\frac{73}{36}-\frac{\pi^2}{6}\right]
\ ,\label{phiLL}
\ee
where $L_0\equiv\ln\phi_0/\mu_0$.
We see that the $r\rightarrow\infty$ ($x\rightarrow\infty$) limit of $\phi_L$
is a constant. Imposing that this constant is $\phi_+=0$, fixes $\phi_0$ to be
\be
\phi_0 = \mu_0\, e^{5/6} \ ,
\label{phi0run}
\ee
confirming the expectation (\ref{phi0exp}) and the previous calculation based on using the Fubini instanton on the constraint (\ref{rescalingrun}). This result makes concrete the generic expectation $\phi_0\sim\mu_0$, which goes back to \cite{Arnold}.

Imposing that the $1/r^2$ terms in both regimes also match gives
\be
C_\phi =\frac{8}{\lambda_0\phi_0}-\frac{2\beta_\lambda'}{3\lambda_0^2\phi_0}\left(\pi^2-\frac{7}{6}\right)\ .
\label{C}
\ee
It can be checked that the first derivatives of $\Phi_{L,R}$ at $r_x$
match, provided $\phi_0$ satisfies (\ref{phi0run}). 

What we have found is that, unlike the case in the previous section, the large $r$ behaviour of the bounce ($\sim 1/r^2$) is not modified by the perturbation, which now simply corrects perturbatively the overall coefficient. In other words, once the condition (\ref{phi0run}) is imposed, the solution (\ref{deltaphiL}) is valid for all $r$. It reads
\be
\delta\phi(r)=\frac{\beta_\lambda'\phi_0}{12\lambda_0(1+x)}\left\{\frac{13}{12}-\frac{1}{x}\ln(1+x)+3\ln^2(1+x)+\frac{x-1}{x+1}\left[\frac{1}{12}+6\mathrm{Li}_2(-x) \right]\right\}\ .
\ee

The tunneling action can then be computed using the previous approximations for the bounce solution and one arrives at
\bea
S_E &= & \frac{8\pi^2}{3\lambda_0}+\frac{2\pi^2\beta_\lambda'}{27\lambda_0^2}(19-30L_0+18L_0^2) +{\cal O}(\beta_\lambda'{}^2)
\label{Sl1}\\
&= & 
\frac{8\pi^2}{3\lambda_0}+\frac{13\pi^2}{27\lambda_0^2}\beta_\lambda' +{\cal O}(\beta_\lambda'{}^2)
\label{Sl2} 
\eea
where $L_0\equiv\ln(\phi_0/\mu_0)$ and we have used $L_0=5/6$ to write the last expression. As expected on general grounds (see discussion in subsection~\ref{subsec:link}) the condition $L_0=5/6$ corresponds to an extremal of the action (\ref{Sl1}).

We can generalize the previous analysis by considering the running to higher loop orders. We write
\be
\lambda(\phi)=-\lambda_0+\int_{\ln\mu_0}^{\ln \phi} \beta_\lambda(\varphi)\, d\ln\varphi\ .
\ee
The usual argument rescaling the bounce or Eq.~(\ref{link}) lead to the condition
\be
\boxed{\int_0^\infty  \phi_B^4 \beta_\lambda(\phi_B)\, r^3dr=0}\ ,
\label{rescalingrungen}
\ee
that generalizes (\ref{rescalingrun}), and was already discussed in \cite{DLIR}, work that is similar to the current paper in exploiting the Affleck condition. The constraint (\ref{rescalingrungen})  shows that a running $\lambda$ is not enough to guarantee a bounce:
$\beta_\lambda$ should change sign for the bounce to exist. 
In this respect, notice that the expansion in
scale-breaking parameters is orthogonal to the loop expansion: in 
(\ref{rescalingrungen}) we could expand $\beta_\lambda$ as a series
of different loop orders but all the terms would be of first order in scale breaking and, in order to satisfy (\ref{rescalingrungen}), the expansion should go at least to two-loops. Finally, as $\mu_0$ is  defined by $\beta_\lambda(\mu_0)=0$ it also follows that the bounce should extend
above $\mu_0$ (that is, $\phi_0>\mu_0$) to fulfill (\ref{rescalingrungen}), confirming that this result is indeed general. 

\subsection{Understanding the result for the action\label{subsec:under}}
To understand further the origin of the result (\ref{Sl1}) notice that
the first-order results for the scaling constraint (\ref{phi0run}) and the action (\ref{Sl1}) can be obtained using the zero-th order Fubini approximation for the bounce profile $\phi_{F}(r)$ of Eq.~(\ref{Fubini}). This allows to change the integration variable in the action and constraint integrals from $r$ to $x\equiv \phi_F/\phi_0$. In this way 
the constraint integral can be rewritten, at first order, as
\be
\int_0^1x(1-x)\beta_\lambda(x\phi_0)dx =0 \ .
\label{Constx}
\ee
Let us define the average of any function $f(\phi)$ over the Fubini bounce with $\phi_F(0)=\phi_0$ by
\be
\langle f(\phi) \rangle_{\phi_0}\equiv 6 \int_0^1x(1-x)f(x\phi_0)dx\ ,
\ee
where the factor 6 is put in so that the average of a constant is the same constant. Using this definition, the bounce constraint (\ref{Constx})
is simply given by
\be
\langle \beta_\lambda\rangle_{\phi_0}=0\ .
\ee
It is immediate to reproduce the result $L_0=5/6$ from this condition when $\beta_\lambda$ is expanded only up to the $\beta_\lambda'$ term, and the result can be extended to higher loop order if needed.

Using the same approach one gets for the tunneling action, at first order,
\be
S\simeq \left\langle\frac{8\pi^2}{3|\lambda(\phi)|}\right\rangle_{\phi_0}\ ,
\ee
understanding that this expression has to be expanded, using 
\be
\lambda(\phi) = -\lambda_0+\int_{\ln\mu_0}^{\ln\phi}\beta_\lambda(\mu)\ d\ln\mu
\ ,
\ee
to get
\be
S\simeq \frac{8\pi^2}{3\lambda_0}\left\langle1+\frac{1}{\lambda_0}\int_{\ln\mu_0}^{\ln\phi}\beta_\lambda(\mu)\ d\ln\mu\right\rangle_{\phi_0}\ .
\ee
Expanding $\beta_\lambda$ up to the $\beta_\lambda'$ term and
 averaging, one immediately reproduces the result (\ref{Sl1}).

\section{The potential  $\bma{V(\phi)=\lambda(\phi)\phi^4/4 }$
with gravity
\label{sec:Vrunlgrav}}

\subsection{Euclidean Action with Gravity}
To include gravitational effects on vacuum decay we follow \cite{CdL}
writing the Euclidean action
\be
S_E =\int d^4 x \sqrt{g}\left[\frac12 Z(\phi)^2g^{\mu\nu}\partial_\mu\phi\partial_\nu\phi+V(\phi)+G(\phi)R\right]+S_{\rm GHY}\ ,
\label{S}
\ee
where we take $Z(\phi)=1$ and
\be
G(\phi)=-\frac{1}{2\kappa}+\frac12\xi\phi^2\ ,
\ee
where $\kappa = 1/m_P^2$, $m_P=2.435\times 10^{18}$ GeV is the reduced Planck mass and a nonminimal coupling $\xi$ of the scalar to gravity is included.
The term $S_{\rm GHY}$ is the Gibbons-Hawking-York boundary term \cite{GHY}, required to get rid of the second-derivatives of the metric and set up a well-posed variational problem in the presence of a boundary, see below.
Assuming $O(4)$-symmetry, we take the Euclidean metric to be
$ds^2= g_{\mu\nu} dx^\mu dx^\nu = dr^2 +\rho(r)^2 d\Omega_3^2$,
where $r$ measures the radial distance along lines normal to three-spheres of radius of curvature
$\rho(r)$ and $d\Omega_3^2$ is the line element on a unit three-sphere.
The Ricci scalar for this metric is
\be
R=\frac{6}{\rho^2}(1-\dot\rho^2-\rho \ddot\rho)\ ,
\label{R}
\ee
where dots stand for derivatives with respect to $r$.
The action (\ref{S}) reads then
\be
S_E=2\pi^2 \int_0^{\infty} dr \left\{\rho^3\left[\frac12 \dot\phi^2+V(\phi)\right]
+6G(\phi)\rho(1-\dot\rho^2-\rho \ddot\rho)\right\}
+\left.12\pi^2 G(\phi)\rho^2\dot\rho\right|_0^{\infty}\ ,
\ee
where the last term is the GHY boundary term. Integrating by parts the $\dot\rho^2$ term removes the $\ddot\rho$, cancels out the GHY term,
and leads to the simpler expression
\be
S_E=2\pi^2 \int_0^{\infty} dr \left\{\rho^3\left[\frac12 \dot\phi^2+V(\phi)\right]
+6\rho\left[G(\phi)(1+\dot\rho^2)+\rho\dot\rho G'(\phi)\dot\phi\right]\right\}\ .
\label{SE}
\ee
The Euler-Lagrange equations following from stationarity of this action under variations of $\phi$ and $\rho$ are,
respectively
\bea
\label{EoMphi}
\ddot\phi +3\frac{\dot\rho}{\rho}\dot\phi  &=&V'+G'R\ ,\\
\dot\rho^2&=&1-\frac{\rho^2}{6G}\left(\frac12\dot\phi^2-V+6\frac{\dot\rho}{\rho}G'\dot\phi\right)\ ,
\label{EoMrho}
\eea
where primes denote derivatives with respect to $\phi$. 
These differential equations should be solved for the bounce solution $\phi_B(r)$ and the metric function $\rho_B(r)$, and in our case, with $V_+\simeq 0$, the boundary conditions are
\be
\phi(\infty)=\phi_+=0\ ,\quad \dot\phi(0)=0\ ,\quad \rho(0)=0\ ,\quad \dot\rho(\infty)=1\ .
\label{BCs}
\ee

The exponential suppression of vacuum decay is controlled by the difference 
$\Delta S_E=S_E[\phi_B]-S_E[\phi_+]$ between the action of the bounce and the action of the false vacuum background $\phi_+=0$, which we take to be well approximated by a Minkowski vacuum $V(\phi_+)\simeq 0$, with the flat metric function $\rho_+(r_+)=r_+$. Substituting these values in (\ref{SE}) one gets
\be
S_E[\phi_+]=24\pi^2 G_E \int_0^\infty dr_+\,  r_+ = 24\pi^2 G_E \int_0^{\infty} dr\,  \rho\dot\rho\ ,
\ee
with $G_E\equiv G(0)=-1/(2\kappa)$. In the last expression we have changed variables to bounce coordinates identifying $\rho_+(r_+)=\rho(r)$. This is convenient in order to write $\Delta S_E$ as a single integral and to enforce the cancellation of divergent contributions in $S_E[\phi_B]$ and $S_E[\phi_+]$ for $r \rightarrow \infty$ so that 
$\Delta S_{E}$ is finite.
In this way one arrives at 
\be
\Delta S_{E}= 2\pi^2\int_0^{\infty}dr \left\{\rho^3\left(\frac12 \dot \phi^2+V\right)+6\rho\left[(1+\dot\rho^2)G-2\dot\rho G_E+\rho\dot\rho
G'\dot \phi\right]\right\}\ .
\label{DSE1}
\ee
%Other simpler expressions can be derived from this by using integration by parts and/or the equations satisfied by $\phi_B$ and $\rho_B$ but we simply stick to (\ref{DSE1}).

\subsection{General Considerations}
The constraint on the bounce from the rescaling argument can be immediately obtained by the method explained in subsection~\ref{subsec:link}. One has
\be
\frac{d\Delta S_{E}}{d\ln\phi_0}[\phi_B,\rho_B]=\int_0^\infty\left(\frac{\delta \Delta S_{E}}{\delta \phi}\frac{\partial \phi_B}{\partial \ln\phi_0}+\frac{\delta \Delta S_{E}}{\delta \rho}\frac{\partial \rho_B}{\partial \ln\phi_0}+\sum_{p_\alpha}\frac{d\Delta s_{E}}{d p_\alpha}\frac{\partial p_\alpha}{\partial \ln\phi_0}\right)dr= 0\ ,
\ee
where $p_\alpha=\{\lambda_0,\beta_\lambda',\mu_0,\kappa,\xi\}$ are the parameters entering the action density
$\Delta s_{E}$. Note that this method has to be used on the action as expressed in (\ref{DSE1}) as this is the action whose variation gives the equations of motion for $\phi$ and $\rho$ so that the first two terms above vanish.
Using (\ref{dims}) and 
\be
\frac{\partial \kappa}{\partial \ln\phi_0}=-2\kappa\ ,\quad
\frac{\partial \xi}{\partial \ln\phi_0}=0 \ ,
\label{dims2}
\ee
one gets
\be
\boxed{
 \int_0^\infty \left[ \rho_B^3\beta_\lambda'\phi_B^4 \ln\frac{\phi_B}{\mu_0}+24m_P^2\rho_B(1-\dot\rho_B)^2\right] dr=0}\ .
\label{rescalingrungrav}
\ee 
This constraint shows explicitly that if $\beta_\lambda'=0$ and only gravity breaks scale invariance  then $\rho_B=r$. This flat metric is possible only for the conformal value $\xi=1/6$ \cite{RS,SSTU}. For generic values of $\xi$ no bounce exists, in agreement with the remarks in \cite{BMZ,RS}.
We then see that the running of $\lambda$ (with varying $\beta_\lambda$) is indeed needed to get the bounce and gravity simply modifies the properties of that bounce (its scale in particular, see below), in agreement with \cite{IRST,SSTU}.

We consider now both $\beta_\lambda'$ and $\kappa=1/m_p^2$ as small perturbations of the scale-invariant problem. Working at first order in these perturbations and writing $\rho_B\simeq r+\kappa \rho_1$, $\phi_B\simeq  \phi_{F}$ with $\phi_F$ the Fubini bounce of Eq.~(\ref{Fubini}), we get
\be
\frac{2\beta_\lambda'}{9\lambda_0^2}\left(\ln\frac{\phi_0}{\mu_0}-\frac{5}{6}\right)+\kappa \int_0^\infty  \dot\rho_1^2 r dr=0\ .
\label{rescalingrungrav}
\ee
As both terms above are positive, we see that gravity reduces the value of $\phi_0$. This effect, shown numerically in Fig.~\ref{fig:phi0rungrav}, is discussed in more detail below.

\subsection{Perturbative Analysis\label{subsec:SMpa}}
We can calculate the ${\cal O}(\kappa)$ corrections to the bounce in the two field regimes already considered in the previous section. In both regimes we write
\be
\phi(r)=\phi_{\kappa 0}(r)+\kappa \phi_1(r) +{\cal O}(\kappa^2)\ ,\quad\, 
\rho(r)=r+\kappa \rho_1(r) +{\cal O}(\kappa^2)\ ,
\ee
where $\phi_{\kappa 0}$ is the bounce without gravity discussed in Section~\ref{sec:Vrunl}. Consider first the small field regime, with $\phi\leq \phi_x$. There
\be
\phi_{\kappa 0}(r)\simeq\phi_S(r)=\frac{C_\phi}{r^2}\ ,
\ee
with $C_\phi$ as given in (\ref{C}). Expanding Eq.~(\ref{EoMrho}) for $\rho$ to order
$\kappa$ we get
\be
24 \xi r \phi_S\dot\phi_S+2r^2\dot\phi_S^2+\lambda_0 r^2\phi_S^4=24\dot\rho_{1S}\ ,
\ee
where $\rho_{1S}$ is the small-field regime approximation to $\rho_1$.
One finds 
\be
\rho_{1S}(r)=C_\rho-\frac{2C_\phi^2}{3r^3}(\xi-1/6)+{\cal O}(1/r^4)\ .
\ee
Expanding Eq.~(\ref{EoMphi}) for the bounce to ${\cal O}(\kappa)$ and
${\cal O}(1/r^4)$ we get
\be
\ddot\phi_{1S}+\frac{3}{r}\dot\phi_{1S}=-3\lambda_0 C_\phi^2\frac{\phi_{1S}}{r^4}\ ,
\ee
which gives $\phi_{1S}= C_{\phi 1}/r^2$. Therefore
\be
\phi_S(r) =\frac{1}{r^2}(C_\phi+\kappa C_{\phi 1})+{\cal O}(\kappa^2)\ . 
\ee

Turning to the large field regime, with $\phi\geq\phi_x$, and solving
the bounce and $\rho$ equations (\ref{EoMphi}) and (\ref{EoMrho}) at linear order in $\kappa$
we get
\be
\rho_{1L}(r)=-(\xi-1/6)\phi_0\sqrt{\frac{2}{\lambda_0}}\left[\frac{x-1}{(x+1)^2}\sqrt{x}+\arctan\sqrt{x}\right]\ ,
\label{rhoL}
\ee
with $x=\lambda_0\phi_0^2r^2/8$, and
\bea
\phi_{1L}(r)&=&-(\xi-1/6)\frac{\phi_0^3}{(1+x^3)}\left\{\frac12\left[x(x-1)-(1+x)\sqrt{x}\arctan(\sqrt{x})\right]\right.\nonumber\\
&+&\left.\frac15(\xi-1/6)\left[6(x^2-1)\ln(1+x)+x(x^2-8x-24)\right]\right\}\ .
\label{phi1L}
\eea
The expected two integration constants are fixed by requiring that $\phi_{1L}(r)$ does not diverge at $r=0$ and that $\phi_{1L}(0)=0$ (as $\phi_0$ at this level is an unknown to be determined, see below). 

We match $\rho_{1S}$ and $\rho_{1L}$ by expanding the latter for
large $r$ and this gives
\be
C_\rho= -(\xi-1/6)\frac{\pi\phi_0}{2\sqrt{2\lambda_0}}\ .
\ee

To match $\phi_{1S}(r)$ and $\phi_{1L}(r)$ we use the large $r$
expansion of the latter, that reads
\be
\phi_{1L}(r)\simeq -\frac15(\xi-1/6)^2\phi_0^3-\frac{6\phi_0^3}{5x}(\xi-1/6)\left[(\xi-1/6)\left(\ln x-\frac{11}{6}\right)+\frac{5}{12}\right]\ .
\ee
Combining this result with the expansion (\ref{phiLL}) we get
\bea
\phi_L(r)+\kappa\phi_{1L}(r)&=&-\left[
\frac{\beta_\lambda'\phi_0}{12\lambda_0} \left(L_0+\frac56\right)-\frac{\kappa}{5}(\xi-1/6)^2\phi_0^3\right]\left(1+\frac{6}{x}\ln x\right)\label{phiLr}\\
&+&\frac{\phi_0}{x}+
\frac{\beta_\lambda'\phi_0}{2\lambda_0 x}\left[
\left(L_0+\frac{11}{6}\right)\ln\frac{\phi_0}{\mu_0}-\frac{73}{36}-\frac{\pi^2}{6}\right]+\kappa\frac{11\phi_0^3}{5x}(\xi-1/6)\left(\xi-\frac{13}{33}\right)\ .
\nonumber
\eea
For $r\rightarrow\infty$ ($x\rightarrow \infty$), $\phi_L+\kappa\phi_{1L}$ above goes to a constant. Imposing that constant to be $\phi_+=0$, leads to the condition 
\be
\boxed{
\frac{\beta_\lambda'}{\lambda_0}\left(\ln\frac{\phi_0}{\mu_0}-\frac{5}{6}\right)+\frac{12}{5}(\xi-1/6)^2\kappa \phi_0^2=0}\ ,
\label{phi0rungrav}
\ee
that fixes $\phi_0$ implicitly. This choice also implies the cancelation of the $(1/x)\ln x$ term in (\ref{phiLr}). 
\begin{figure}[t!]
\begin{center}
\includegraphics[width=\textwidth]{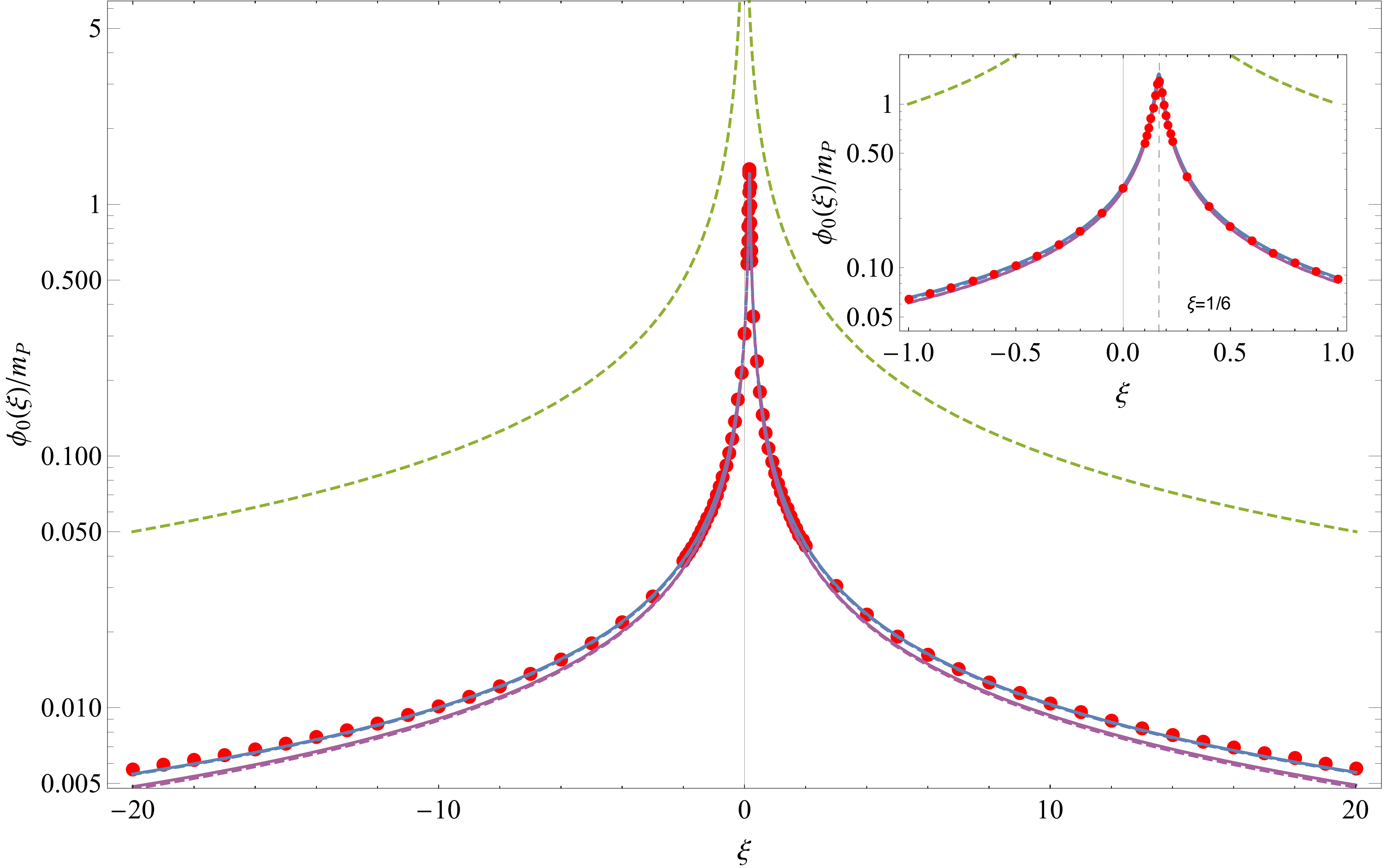}
\end{center}
\caption{Ratio $\phi_0/m_P$ as a function of the non-minimal coupling $\xi$ for the SM with $M_h=125.09$ GeV and $M_t=173.34$ GeV. Red points: from full numerical solutions for bounce and metric function. Violet lines from analytical approximations:  dashed from  Eq.~(\ref{phi0rungrav})  and solid from Eq.~(\ref{Const2}). Blue lines from improved analytical approximations:  dashed from  Eq.~(\ref{Consa})  and solid from Eq.~(\ref{Cons2a}).
The maximum corresponds to the conformal value $\xi=1/6$ (see inset).
The dashed line gives the condition $\phi_0/m_P<\Lambda_{UV}/m_P=1/|\xi|$.
\label{fig:phi0rungrav}
}
\end{figure}
We get the same result (\ref{phi0rungrav}) if we plug $\rho_1$ in (\ref{rescalingrungrav}) and perform the integral. Fig.~\ref{fig:phi0rungrav} shows with a dashed violet line  the numerical solution of (\ref{phi0rungrav}) for $\phi_0/m_P$ as a function of $\xi$ with $\lambda_0=0.0143$, $\beta_\lambda'=5.6\times 10^{-5}$ and $\mu_0/m_P=0.59$, values that correspond to the central values $M_h=125.09$ GeV and $M_t=173.34$ GeV in the SM. The red points in the same plot correspond to numerically solving the differential equations (\ref{EoMphi}) and (\ref{EoMrho}) for the bounce and the metric function for the same choice of parameters. We used the NNLO Higgs effective potential, calculated at two-loops, with parameters running at three loops and two-loop matching conditions \cite{SMnew}. The agreement between (\ref{phi0rungrav}) and the full numerical solution is quite good for low $|\xi|$ getting worst for larger values: the error is $\sim 16\%$ at $|\xi|\sim 20$.

Some comments on the result (\ref{phi0rungrav}) are in order:

(1) As is well known, the inclusion of gravity makes the theory non-renormalizable. When $\xi^2\gg 1$ the cutoff of the effective theory is $\Lambda_{UV}=m_P/|\xi|$ \cite{BLT,BE}. The gravitational contribution to (\ref{phi0rungrav}) reflects this through the 
appearance of the mass ratio $\phi_0/\Lambda_{UV}$.
To have control over the tunneling calculation one needs to impose $\phi_0\lsim m_P/\xi$ and Fig.~\ref{fig:phi0rungrav} shows that $\phi_0$ is always safely below $\Lambda_{UV}$. On the other hand, this condition implies $(\xi-1/6)^2\kappa\phi^2\lsim 1$
so that the gravitational correction to the tunneling action cannot be as important as the non-gravitational one\footnote{This assumes that the potential is well approximated by $V=-\lambda\phi^4/4$ with a nearly constant $\lambda$, an assumption that fails near the critical Higgs mass case, for which the vacuum at high field values is almost degenerate with the electroweak one. In that case, gravitational corrections can prevent vacuum decay, as is well known in general \cite{CdL}. (See \cite{EFT} for an analysis of this case in the SM.)}. 

(2) Fig.~\ref{fig:phi0rungrav} shows that, for large values of $|\xi|$, the bounce field value $\phi_0$ can be significantly lower than $m_P$, although it is gravity that causes the decrease. The reason is, once again, that the relevant scale to determine gravitational effects is $\Lambda_{UV}=m_P/|\xi|$ rather than $m_P$ and typically $\phi_0$ is just one order of magnitude below $\Lambda_{UV}$.\footnote{This has a bearing on the claim \cite{Brxi} that a nonmiminal coupling reduces the impact of nonrenormalizable operators suppressed by $m_P$ on the stability of the potential, which can be understood from $\phi_0\ll m_P$. However, for nonzero $\xi$, the dominant operators are suppressed by $m_P/\xi$ rather than $m_P$. }

As in the case without gravity we see that, once the condition that determines $\phi_0$ is imposed, the solutions (\ref{rhoL}) and (\ref{phi1L}) are valid for all $r$. One gets ($x=\lambda_0\phi_0^2r^2/8$)
\bea
\delta\phi(r)&=&
\frac{\beta_\lambda'\phi_0}{12\lambda_0(1+x)}\left\{\frac{13}{12}-\frac{1}{x}\ln(1+x)+3\ln^2(1+x)+\frac{x-1}{x+1}\left[\frac{1}{12}+6\mathrm{Li}_2(-x) \right]\right\}\nonumber\\
&+&\kappa\frac{(\xi-1/6)x \phi_0^3}{2(1+x)^2}\left\{\frac{1-x}{1+x}+\frac{1}{\sqrt{x}}\arctan\sqrt{x}
+2(\xi-1/6)\left[\frac{1-2x}{1+x}+\frac{12}{5x}\ln(1+x)\right]
\right\}\nonumber\\
&+&\frac{72\kappa^2}{25\beta_\lambda'}\frac{\lambda_0(\xi-1/6)^4x\phi_0^5}{(1+x)^2}\ ,
\label{phi1}
\eea
and $\rho_1(r)$ as given in (\ref{rhoL}).

We now have all the ingredients to calculate the Euclidean tunneling
action. We get
\bea
S_E&\simeq&\frac{8 \pi^{2}}{3 \lambda_0}+\frac{2 \pi^{2} \beta_\lambda'}{27 \lambda_0^{2}}\left(19-30 L_0+18 L_0^{2}\right)+\frac{16 \pi^{2} }{5 \lambda_0}  (\xi-1 / 6)^{2} \kappa\phi_0^{2}
\label{Sg1}\\
&\simeq &
\frac{8\pi^2}{3\lambda_0}+\frac{13\pi^2}{27\lambda_0^2}\beta_\lambda'+\frac{16\pi^2}{5\lambda_0}(\xi-1/6)^2\kappa \phi_0^2
+\frac{192\pi^2}{25\beta_\lambda'}(\xi-1/6)^4\kappa^2\phi_0^4\ .
\label{Sg2}\eea
In this formula, $\phi_0$ is the solution of (\ref{phi0rungrav}), which we have used to write the last expression. As in all the examples in previous sections, the condition (\ref{phi0rungrav}) corresponds in fact to $dS_E/d\phi_0=0$, with $S_E$ as given in (\ref{Sg1}).

\begin{figure}[t!]
\begin{center}
\includegraphics[width=\textwidth]{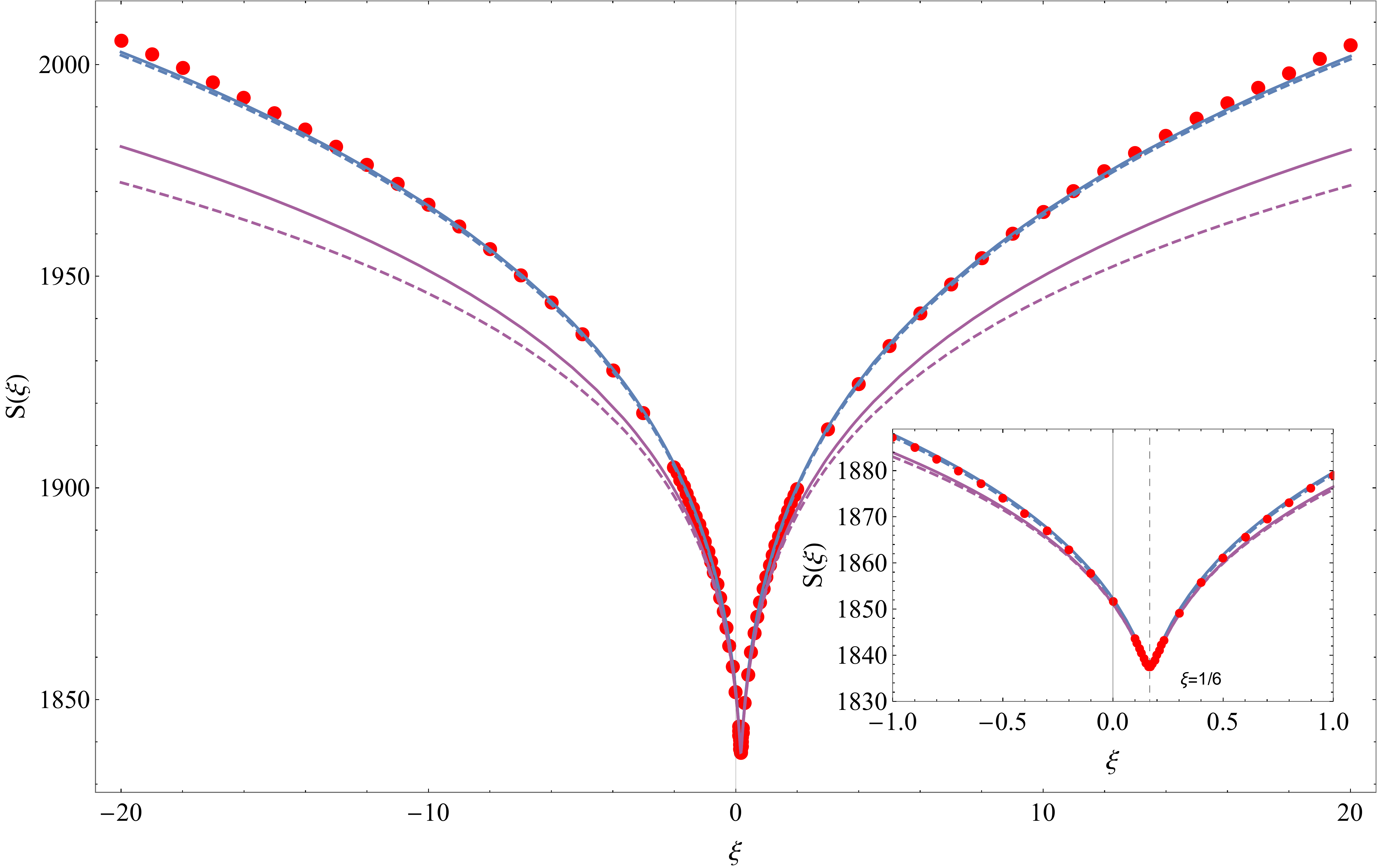}
\end{center}
\caption{Tunneling action as a function of the non-minimal coupling $\xi$ for the SM with $M_h=125.09$ GeV and $M_t=173.34$ GeV. 
Red points: from full numerical solutions for bounce and metric function. Violet lines from analytical approximations:  dashed from  Eq.~(\ref{Sg1})  and solid from Eq.~(\ref{S2Vt}). Blue lines from improved analytical approximations:  dashed from  Eq.~(\ref{S1a})  and solid from Eq.~(\ref{S2a}).
The minimum corresponds to the conformal value $\xi=1/6$.
\label{fig:Sxi}
}
\end{figure} 

Some comments on the result (\ref{Sg1}) are in order:

(1) The gravitational correction vanishes for the conformal value $\xi=1/6$, as is well known. However, once the running effects are
included, the potential is not scale invariant and at higher orders in
the perturbative expansion gravitational corrections can appear even for
$\xi=1/6$.

(2) If we compare (\ref{Sg1}) with the results of \cite{SSTU} we find the same gravitational correction. Nevertheless, to get the final analytic result for the action it is crucial to use the constraint (\ref{phi0rungrav}) that links the gravitational piece with the one due to the running of $\lambda$, which was not made explicit in \cite{SSTU}. Comparison of (\ref{Sl2}) and (\ref{Sg2}) shows that it would be wrong to simply add the gravitational piece directly to (\ref{Sl2}).

(3) In \cite{IRST,SSTU} the Euclidean action was expressed in terms of $\rho_1$ only, avoiding the need to calculate $\phi_1$, and the theoretical approximation to the action in \cite{SSTU} was not used directly but rather evaluating numerically the Euclidean action over the Fubini instanton (with the running coupling evaluated on the Fubini instanton too).  Not having $\phi_1$, however, prevents going to higher order in the perturbative expansion. Our approach puts on a firmer basis this kind of perturbative analysis, offers an explicit analytic result (having $\phi_1$ under theoretical control), and allows to calculate  higher order corrections.

Fig.~\ref{fig:Sxi} shows with a dashed violet line the analytical approximation (\ref{Sg1})
for the tunneling action compared with the full numerical result obtained by solving for the bounce and metric (red points) for the SM parameters $M_h=125.09$ GeV and $M_t=173.34$ GeV.
The agreement is good although the error grows with $|\xi|$ and is of order
$1.7\%$ for $|\xi|\sim 20$. This error is much smaller than the one for $\phi_0/m_P$ as the action is an extremal at
$\phi_0$, and can be lowered significantly as follows.

It is possible to go beyond the fixed order expansions we have derived so far
to improve the agreement with the fully numerical results. This is especially relevant at large $\xi$ when $\phi_0$ can be orders of magnitude lower than $\mu_0$ so that expanding around $\mu_0$ is not optimal. It is straightforward to rewrite the  constraint (\ref{phi0rungrav}) and action (\ref{Sg1})  without reference to $\mu_0$ in terms of $\lambda$ and $\beta_\lambda$ evaluated at a scale $\phi_0/a$, where $a\sim {\cal O}(1)$ can be varied to optimize agreement. Different choices of the scale at which to evaluate $\beta_\lambda'$ impact the tunneling action by higher order corrections. Still we find it is better to take $\beta_\lambda'=\beta_\lambda'(\phi_0/10)$ and we do that in the plots. Writing
\be
\lambda_a=-\lambda(\phi_0/a)\ ,\quad
\beta_{\lambda,a}=\beta_\lambda(\phi_0/a)\ ,
\ee
the constraint equation for $\phi_0$ takes the form
\be
0=\beta_{\lambda,a} +\beta_{\lambda}' \left(L_a -\frac56 \right)+
\frac{12}{5}\lambda_a (\xi-1/6)^2\kappa \phi_0^2\ .
\label{Consa}
\ee
Using the same quantities, the tunneling action reads
\be
S=
\frac{8 \pi^{2}}{3 \lambda_a}+\frac{8 \pi^{2} }{3 \lambda_a^{2}}\left[\frac12\left(\frac{19}{18}-L_a^2\right)\beta_\lambda'+\left(L_a-\frac56\right) \beta_{\lambda,a}\right]+\frac{16 \pi^{2} }{5 \lambda_a}  y_\xi^{2} \kappa\phi_0^{2}\ .
\label{S1a}
\ee
These expansions are much closer to the fully numerical results as is shown in Fig.~\ref{fig:phi0rungrav} and Fig.~\ref{fig:Sxi} by the blue dashed lines. 
For $|\xi|\sim 20$ the error in $\phi_0/m_P$ is down to $4.4\%$
while for $S(\xi)$ is $0.17\%$. In these figures, we choose $a=2$, value that is well justified by the discussion in Subsection~\ref{subsec:under}.

\section{Tunneling Potential Approach\label{sec:Vt}}

In this section we reconsider the potential of the previous section (SM potential with running quartic and gravity) using a novel approach for the calculation of tunneling actions. This new method has some advantages in terms of simplicity and, after reproducing the results for the action derived in the previous section using the Euclidean approach, we use it in the next section to extend the calculation to second order in perturbations.

The alternative method was first introduced in \cite{E} and 
was later on extended to include gravitational corrections in \cite{Eg}. This approach recasts 
the original problem into an elementary variational problem: find the ``tunneling potential''
$V_t(\phi)$ that interpolates between the symmetric false vacuum
at $\phi_+$  and the
basin of the true vacuum phase and minimizes an action functional, an integral 
in field space of the appropriate action density, that takes the simple form:
\be
 S[V_t]=6\pi^2m_P^4\int_{\phi_+}^{\phi_0}d\phi\ \frac{(D+V_t')^2}{V_t^2 D}\ ,
\ee
 where $D^2\equiv V_t'{}^2+6\kappa (V-V_t)V_t$ and $\phi_0$ is in the basin of the true vacuum.
 
\subsection{Extension to Nonminimal Coupling} 
To apply this new approach to the SM vacuum decay in the presence of a nonminimal coupling of the Higgs to the Ricci scalar we need to extend further the formalism (as \cite{Eg} implicitly assumed $\xi=0$). For the Euclidean action in (\ref{S}) one can obtain the corresponding $S[V_t]$ either by repeating  the procedure used in
\cite{Eg} (done for the special case $Z=1$ and $\xi=0$, {\it i.e.} an Einstein-frame action) or by using a Weyl transformation 
to the Jordan frame with action (\ref{S}).
The end result is \cite{Estab}
\be
S[V_t]=24\pi^2\int_{\phi_+}^{\phi_0}d\phi\ G^2\frac{(D+\hat V_t')^2}{V_t^2 D}\ ,
\label{Sgen}
\ee
with
\be
D  \equiv \sqrt{\hat{V}_t'{}^2-\frac{3}{G}\hat Z^2 (V-V_t)V_t}\ .
\ee
We have introduced the combinations
\be
\hat{V}_t'\equiv V_t'-\frac{2 V_t G'}{G}\ , \quad
\hat Z^2\equiv Z^2-\frac{3G'{}^2}{G}\ .
\ee
As in the previous section we are interested in the case
$Z(\phi)=1$ and $G(\phi)=(-1/\kappa+\xi\phi^2)/2$. 

The Euler-Lagrange equation following from the stationarity of the action (\ref{Sgen}) under
variations of $V_t$ is:
\be
2(V-V_t)\left[\hat V_t''-\left(\frac{3G'}{2G}+\frac{\hat Z'}{\hat Z}\right)\hat V_t'-\frac{\hat Z^2}{G}\left(
\frac{3}{2}V-V_t\right)\right]
+\hat V_t'\left[\frac43 \hat V_t'-V'+2V\frac{G'}{G}\right]=0
\ .
\label{VtEoMg}
\ee
In the case of the false vacuum being a Minkowski vacuum (or nearly so
as for the SM potential) the boundary conditions for $V_t$
are  
\cite{Eg}
\be
V_t(\phi_+)=V(\phi_+)=0\ ,\quad\quad  V_t(\phi_0)=V(\phi_0)\ ,
\label{BCsVt}
\ee
with $\phi_0$ [equal to $\phi_B(0)$ in the Euclidean bounce approach] being an unknown to be found.
The equation of motion for $V_t$ also fixes the derivatives at the two extremes of the interval $(\phi_+,\phi_0)$ as
\be
V_t'(\phi_+)=V'(\phi_+)=0\ ,\quad
V_t'(\phi_0)=\frac34 V'(\phi_0)+\frac12 V(\phi_0)\frac{G'(\phi_0)}{G(\phi_0)}\ .
\label{BCsVtp}
\ee

\subsection{Perturbative Analysis}
To understand the impact of gravity on the decay of Minkowski (or AdS) vacua it is instructive to expand the tunneling potential, the Euler-Lagrange equation for $V_t$ and the tunneling action to first order in $\kappa$. We write 
\be
V_t=V_{t\kappa 0}+\kappa V_{t\kappa 1}+{\cal O}(\kappa^2)\ ,
\ee 
where $V_{t\kappa 0}$ is the tunneling potential without gravity. 
The zeroth-order EoM for $V_{t\kappa 0}$ is
\be
\label{EoM0}
(4V_{t\kappa 0}'-3V')V_{t\kappa 0}'=6(V_{t\kappa 0}-V)V_{t\kappa 0}''\ .
\ee
The general expansion of the tunneling action density ${\it s}(\phi)$ can be written simply as \cite{Estab}
\be
{\it s}(\phi)={\it s}_{\kappa 0}(\phi)\left[1+3\kappa 
\left(2 \xi \phi -\frac{V_{t\kappa 0}}{V_{t\kappa 0}'}\right)^2\right]
+{\cal O}(\kappa^2)\ ,
\label{sexp}
\ee
up to a total-derivative term discussed below.
The zero-th order term in this expansion reproduces the tunneling action density
in the absence of gravity \cite{E}
\be
{\it s}_{\kappa 0}(\phi)=-54\pi^2\frac{(V-V_{t\kappa 0})^2}{V_{t\kappa 0}'^3}\ .
\ee
As $V'_{t\kappa 0}\leq 0$ \cite{E}, this contribution is always positive. The second term in (\ref{sexp}) gives the ${\cal O}(\kappa)$ effects of gravity on the tunneling action and is also positive definite.  So we see from here that gravity makes metastable vacua more stable.\footnote{
Besides proving that gravity makes the action density larger one should care about how gravity changes the integration interval via changes in the end-point of that interval, $\phi_0$. However, an ${\cal O}(\kappa)$ change in $\phi_0$ does not modify the action integral at that order as ${\it s}(\phi_0)=0$.} Moreover, turning on a large $\xi$ can add further to the stabilization effect. Although this perturbative proof holds at ${\cal O}(\kappa)$ the stabilizing effect of gravity (for Minkowski or AdS vacua)
can be proven in more generality \cite{Estab}. 

We omitted from (\ref{sexp}) a term that after using the EoM for $V_t$ is a total derivative that contributes a boundary term to the integrated tunneling action
\be
\delta {\it s}(\phi)=
162\pi^2\kappa\frac{d}{d\phi}\left\{\frac{(V-V_{t\kappa 0})^2}{V_{t\kappa 0}'^4}\left[V_{t\kappa 1}+\frac{(V-V_{t\kappa 0})V_{t\kappa 0}^2}{V_{t\kappa 0}'^2}\right] \right\}\ .
\label{sexpb}
\ee
However this boundary term does not contribute to the action integral: at $\phi_0$ one has $V_{t\kappa 0}(\phi_0)=V(\phi_0)$ with
non-zero $V'_{t\kappa 0}(\phi_0)$ so that the boundary term vanishes at $\phi_0$. To show that the boundary term also vanishes at $\phi_+$ one needs to know how the functions $V, V_{t\kappa 0}$ and $V_{t\kappa 1}$ approach zero.
The general proof is given in \cite{Estab}. For the current potential it is enough to know that  for $\phi\rightarrow 0$ we have $V\sim {\cal O}(\phi^4)$, $V_t\sim {\cal O}(\phi^3)$, see below. 

An interesting property of the expansion result (\ref{sexp}) is that the first order  ${\cal O}(\kappa)$ term depends only on zero-th order quantities. This is ultimately due to the fact that we are expanding the action around its minimum so that a first order shift in $V_t$ affects the action value only at second order. This is exploited in the next subsection in applying this method to the SM case.

\subsection{Application to the Standard Model}
As in the previous sections, let us expand the tunneling potential  in powers of $\beta_\lambda'$ and $\kappa$ [taken to be of ${\cal O}(\epsilon_0)$] writing
\be
V_t(\phi)= V_{t,0}(\phi) + V_{t,1}(\phi)+{\cal O}(\epsilon_0^2)\ ,
\ee
where $V_{t,1}\sim{\cal O}(\epsilon_0)$.  For the potential we similarly write
\be
V=V_0+ V_1 = -\frac14\lambda_0\phi^4 +\frac18\beta_\lambda'\phi^4\ln^2\frac{\phi}{\mu_0} \ .
\ee
At zero-th order, without running or gravitational effects, the EoM for $V_{t,0}$ is 
\be
6(V_0-V_{t,0})V_{t,0}''+V_{t,0}'(4V_{t,0}'-3V_0') = 0\ ,
\ee
with boundary conditions  
\be
V_{t,0}(\phi_+)=0\ ,\quad  V_{t,0}(\phi_0)=V_0(\phi_0)\ ,\quad
V_{t,0}'(\phi_+)=0\ ,\quad
V_{t,0}'(\phi_0)=\frac34 V_0'(\phi_0)\ ,
\ee
and the action 
\be
S[V_t]=
-54\pi^2\int_{\phi_+}^{\phi_0}\frac{(V-V_{t,0})^2}{V_{t,0}'{}^3}d\phi+{\cal O}(\epsilon_0)
\label{SVtk0}
\ee

The solution is quite simple \cite{E}:
\be
V_{t,0}(\phi) = -\frac{\lambda_0}{4}\phi_0\phi^3\ ,
\ee
with  arbitrary $\phi_0$, leading to
\be
S[V_t]=\frac{8\pi^2}{3\lambda_0} +{\cal O}(\epsilon_0)\ .
\ee

Interestingly, we do not need to solve the EoM for $V_{t,1}$ in order to get the tunneling action at ${\cal O}(\epsilon_0)$. We saw this in the previous subsection for the  ${\cal O}(\kappa)$ corrections and the same holds for the ${\cal O}({\beta'_\lambda})$ ones. To see this latter point, we just need to consider ${\cal O}({\beta'_\lambda})$ perturbations of the action without gravity (\ref{SVtk0}). We only care about the correction $V_1$ to the potential and can ignore the correction from $V_{t,1}$ because the action functional is stationary with respect to changes of $V_t$. We then simply have 
\be
\delta s(\phi) = -108\pi^2\frac{(V_0-V_{t,0})}{V'_{t,0}{}^3}V_1\ .
\ee
Putting this together with the results (\ref{sexp}) and (\ref{sexpb}), the
expansion of the action density up to ${\cal O}(\epsilon_0)$ is simply
\bea
{\it s}(\phi)&=&\frac{8\pi^2}{\lambda_0\phi_0^3}(\phi-\phi_0)^2\left[1+12\kappa\phi^2(\xi-1/6)^2\right]-\frac{8\pi^2}{\lambda_0^2\phi_0^4}(\phi-\phi_0)\phi\beta_\lambda'\ln^2\frac{\phi}{\mu_0}\nonumber\\
&+&\frac{32\pi^2}{\lambda_0^2\phi_0^4}\frac{d}{d\phi}\left[\frac{(\phi-\phi_0)^2}{\phi^2}V_{t,1}(\phi)-\frac{9}{4}\lambda_0\kappa\phi^3(\phi-\phi_0)^3\right]\ .
\eea
The last term is a total derivative and contributes a boundary term to the action that vanishes, due to the fact that $V_{t,1}$ must go to zero as $V_{t,0}\sim \phi^3$.
The rest of the terms can be integrated immediately to get
\be
S=\int_0^{\phi_0}{\it s}(\phi)=\frac{8\pi^2}{3\lambda_0}+\frac{2\pi^2}{27\lambda_0^2}\beta_\lambda'(18L_0^2-30L_0+19)+\frac{16\pi^2}{5\lambda_0}(\xi-1/6)^2\kappa\phi_0^2\ ,
\label{SVt}
\ee
where $L_0\equiv\ln(\phi_0/\mu_0)$. This reproduces in a simple way the result (\ref{Sg1}) found using the Euclidean bounce approach.

For completeness, and for later use, one can solve the EoM for $V_{t1}$, which reads
\bea
0&=&6(V_0-V_{t,0})\left[V_{t,1}''+\kappa (3V_0-2V_{t,0}+4\xi V_{t,0}-6\xi^2\phi V_{t,0}')\right]
+(8 V_{t,0}'-3V_0')V_{t,1}'\nonumber\\
&-&3V_1'V_{t,0}'+6(V_1-V_{t,1})V_{t,0}''
+ 2\kappa \xi\phi\left[5(3V_0-V_{t,0})V_{t,0}'-6V_{t,0}V_0'\right]\ ,
\label{EoMVt1}
\eea
with boundary conditions
\be
V_{t,1}(\phi_+)=0\ ,\quad  V_{t,1}(\phi_0)=\frac{\phi_0^4}{8}\beta_\lambda'\ln^2\frac{\phi_0}{\mu_0} ,\quad
V_{t,1}'(\phi_+)=0\ ,\quad
V_{t,1}'(\phi_0)=-\kappa\xi\phi_0 V_0(\phi_0)\ ,
\label{BCs1}
\ee
getting
\bea
V_{t1}(\phi)&=& C_1 \phi^3+C_2 \phi^2\left[2\phi_0+\frac{\phi_0^2}{\phi-\phi_0}+2\phi\ln\left(\frac{\phi_0}{\phi}-1\right)\right]+
\frac{\beta_\lambda'\phi^3\phi_0}{8}\left[\frac{37}{36}+\mathrm{Li}_2\left(\frac{\phi}{\phi_0}\right)\right]\nonumber\\
&+&\lambda_0\kappa\phi^3\left[\frac{\phi^2}{24}(\phi-\phi_0)+\frac{\phi_0}{8}(\phi^2-\phi_0^2)(\xi-1/6)-\frac{3\phi_0}{20}(3\phi^2-3\phi_0\phi+\phi_0^2)(\xi-1/6)^2
\right]\nonumber\\
&+&
\frac{\phi^3\phi_0}{16}\left[
\frac{\phi}{\phi-\phi_0}+2\ln\left(1-\frac{\phi}{\phi_0}\right)\right]\left[\beta_\lambda'\left(\ln\frac{\phi}{\mu_0}-\frac56\right)+\frac{12}{5}(\xi-1/6)^2\lambda_0\kappa\phi_0^2\right] ,
\label{Vt10}
\eea
where $C_{1,2}$ are integration constants. This expression contains divergences in the limit $\phi\rightarrow\phi_0$ that require 
\be
C_2=-\frac{\phi_0}{16}\left[\beta_\lambda'\left(\ln\frac{\phi_0}{\mu_0}-\frac56\right)+\frac{12}{5}\lambda_0(\xi-1/6)^2\kappa\phi_0^2\right]=0\ ,
\ee
the last equality following from (\ref{phi0rungrav}).\footnote{Alternatively, a small field expansion gives
$V_{t1}(\phi)\simeq C_2\phi_0\phi^2+{\cal O}(\phi^3)$ which would imply that the Euclidean bounce does not extend to $r=\infty$ \cite{En}, which again shows $C_2=0$ is needed.} The $\phi\rightarrow\phi_0$ divergences coming from the last line of (\ref{Vt10})
are also absent as the last bracket vanishes for $\phi=\phi_0$, precisely due to  the same condition  (\ref{phi0rungrav}).
Finally, $C_1$ is fixed to
\be
C_1=-\frac{1}{144}\phi_0\beta_\lambda'\left(
20+9L_0-18L_0^2+3\pi^2\right)\ ,
\ee
by demanding $V_{t}(\phi_0)=V(\phi_0)$. With all these ingredients we arrive at the final form
\bea
V_t(\phi)&=&-\frac{1}{4}\lambda_0\phi^3\phi_0\left\{1-\frac{\beta_\lambda'}{4\lambda_0}\left[\frac{7}{18}+\frac{x\ln x}{x-1}-2\mathrm{Li}_2\left(1-x\right)\right]\right.\label{Vt1}\\
&-&\left.\frac{1}{6}\kappa\phi_0^2\left[
x^2(x-1)+3 (x^2-1)(\xi-1/6)-\frac65 (\xi-1/6)^2\left[6L_0+5+9 x(x-1) \right]\right]\right\}\ ,\nonumber
\eea
where now $x\equiv\phi/\phi_0$.
It can be checked that using this $V_t$ in the action formula  (\ref{Sgen}) reproduces the result (\ref{SVt}). In terms of simplicity, the single result for $V_t$ in (\ref{Vt1}) should be compared with the expressions for $\rho_{1}(r)$ in (\ref{rhoL}) and  $\phi_1(r)$ in (\ref{phi1}) for the Euclidean bounce approach.

\section{Second Order Corrections\label{sec:high}}

It is straightforward to add the ${\cal O}(\epsilon_0^2)$ corrections to the results 
for the integral constraint and the tunneling action obtained in the previous section. The scale-breaking condition that fixes $\phi_0$
can be obtained from the general formula (\ref{rescalingrungrav})
expanded to ${\cal O}(\epsilon_0^2)$. It depends on $\rho_2'(r)$,
that can be eliminated in terms of ${\cal O}(\epsilon_0)$ quantities
from  equation (\ref{EoMrho}), so that only $h_0(r)$, $h_1(r)$
and $\rho_1(r)$ are needed to obtain that condition for $\phi_0$ to ${\cal O}(\epsilon_0^2)$. It reads
\bea
0&=&\beta_\lambda'(L_0-5/6)+\frac{12}{5}\lambda_0(\xi-1/6)^2\kappa \phi_0^2+
\frac{(\beta_\lambda')^2}{12\lambda_0}\left(12L_0^3-18L_0^2+19L_0-17+\pi^2\right)
\nonumber\\
&+&\frac{1}{25}\beta_\lambda'\kappa\phi_0^2(\xi-1/6)\left[\frac{97}{8}-\frac{15}{2}L_0+(\xi-1/6)\left(30L_0^2+101L_0+10\pi^2-\frac{5777}{60}\right)\right]
\nonumber\\
&+&
\frac{12}{35}\lambda_0\kappa^2\phi_0^4(\xi-1/6)^3\left[\frac{809}{25}(\xi-1/6)-1\right]\ .
\label{Const2}
\eea
The same result can be obtained in the tunneling potential approach. The general expression for the constraint integral in that formulation can be derived simply by expanding the action integral to ${\cal O}(\epsilon_0^2)$ and then equating to zero its derivative with respect to $\log\phi_0$. Formally the constraint takes the form
\be
0=\int_{\phi_+}^{\phi_0}\left[\frac{\partial s(\phi)}{\partial \kappa}(-2\kappa)+
\frac{\partial s(\phi)}{\partial V}\frac{\partial V}{\partial \mu_0}\mu_0\right] d\phi\ ,
\label{condVt}
\ee
where we have used (\ref{dims}) and (\ref{dims2}).\footnote{A similar formula can be derived for any other potential that depends on parameters $p_\alpha$, with the derivative $(\partial V/\partial\mu_0)$ in (\ref{condVt}) replaced by $\sum_\alpha(\partial V/\partial p_\alpha)d_{(\alpha)} p_\alpha$, where $d_{(\alpha)}$ is the engineering dimension of $p_\alpha$.}

As the tunneling potential method is in fact somewhat simpler for calculating the tunneling action, it is the one we detail here (nevertheless the result below has been cross-checked following the Euclidean method also). Explicitly one gets for the ${\cal O}(\epsilon_0^2)$ correction to the action density:
\bea
s_2(\phi)&=&\frac{32 \pi^2}{\lambda_0^2}
\frac{d}{d\phi}\left\{(x-1)\left[\frac{(x-1)}{x^2}v_{t2}(x)
-\frac29\kappa\phi_0^2(x-1)\left(5x-\frac52+6y_\xi-54y_\xi^2\right) v_{t1}(x)
 \right.\right.\nonumber\\
 &-&\left.\left.\frac{\beta_\lambda'}{\lambda_0 x} v_{t1}(x)L_\phi^2+\frac{4}{\lambda_0 x^5} v_{t1}^2(x)+g(x) \right]\right\} +\frac{2\pi^2x^2}{\lambda_0 \phi_0}\left[ \frac{4\beta_\lambda'}{3\lambda_0} 
\kappa\phi_0^2 x(x - 1) \left(3x - 2-36 y_\xi^2\right) L_\phi^2\right.\nonumber\\
 &+& \left.\frac{\beta_\lambda'{}^2}{\lambda_0^2}L_\phi^4 +\frac17\kappa^2\phi_0^4 x^2  \left(\frac{1}{162} - 
    \frac{1}{27} y_\xi - \frac{10}{9} y_\xi^2 -16 y_\xi^3+48y_\xi^4\right)\right]\nonumber\\
&+&\frac{64 \pi^2(x-1)}{\lambda_0^2 x \phi_0}
\left\{\frac{\beta_\lambda'}{\lambda_0 x} v_{t1}(x) L_\phi+\frac{8}{\lambda_0 x^5} v_{t1}^2(x) +  \frac{4 (x - 1)}{3\lambda_0 x^3}v_{t1}'{}^2(x) \right.\nonumber\\
&+&\left.\frac23 \kappa\phi_0^2\left[1-\frac{11x}{3}+3x^2+(5x-3)y_\xi+18(1-x)y_\xi^2\right]v_{t1}(x)\right\}\ ,
\label{s2Vt}
\eea
where $x\equiv\phi/\phi_0$, $y_\xi=\xi-1/6$, and
\be
v_{t1}(x)\equiv \frac{V_{t1}(\phi)}{\phi_0^4}\ ,\quad
v_{t2}(x)\equiv \frac{V_{t2}(\phi)}{\phi_0^4}\ ,\quad
v'_{t1}(x)\equiv \frac{V'_{t1}(\phi)}{\phi_0^3}\ , \quad
L_\phi\equiv\ln\frac{\phi}{\mu_0}\ .
\ee
The function $g(x)$ is simply a polynomial in $x$ with $g(0)=0$. We do not write it explicitly as this is the only property that matters. Upon integration, the total derivative term in (\ref{s2Vt}) gives a boundary contribution that vanishes: the function inside the curly brackets is zero at $x=0$ and $x=1$. This is guaranteed by the fact that for $\phi\rightarrow 0$, $V_{ti}(\phi)\sim\phi^3$.
This means that $V_{t2}(\phi)$ is not needed to calculate the tunneling action at ${\cal O}(\epsilon_0^2)$ and it suffices to know $V_{t1}(\phi)$,
which is given in the previous section. Using that result and performing the integral one arrives at
\bea
S&=&
\frac{8 \pi^{2}}{3 \lambda_0}+\frac{2 \pi^{2} \beta_\lambda'}{27 \lambda_0^{2}}\left(19-30 L_0+18 L_0^{2}\right)+\frac{16 \pi^{2} }{5 \lambda_0}  (\xi-1 / 6)^{2} \kappa\phi_0^{2}\nonumber\\
&+& \frac{\pi^{2} \beta_\lambda'^{2}}{9 \lambda_0^{3}}\left[18 K_0+\frac{553}{36}-\frac{11\pi^{2}}{6}+2\left(-12+\pi^{2}\right) L_0+8 L_0^{2}-8 L_0^{3}+6 L_0^{4}\right] \nonumber\\
&+& \frac{\pi^{2} \beta_\lambda' }{5 \lambda_0^{2}}(\xi-1/ 6)\kappa\phi_0^{2}\left[-\frac{419}{90}+\frac{14}{3}L_0+\left(\frac{8 \pi^{2}}{3}-\frac{623}{25}-\frac{132}{5}L_0+40L_0^2\right)
(\xi-1 / 6)\right]
 \nonumber\\
 &+& \frac{8\pi^{2}}{35 \lambda_0}(\xi-1 / 6)^{3}\kappa^2 \phi_0^{4} \left[6+\frac{1}{25}\left(739+840L_0\right)(\xi-1/6) \right]\ ,
\label{S2Vt}
\eea
where $L_0=\ln\phi_0/\mu_0$ and
\be
K_0=\int_0^1(1-6x+6x^2)\left[\mathrm{Li}_2(x)\right]^2\, dx\ \simeq 0.118718\ .
\ee
Some comments on the result for $S$ above are in order:

(1) For large $|\xi|$, the powers of the potentially large $\xi$ contributions to $S$ above are limited by the fact that whenever $\kappa\phi_0^2\xi^n$ appears, one always has $n\leq 2$ as necessary to be consistent with $\Lambda_{UV}=m_P/|\xi|$.
Of course, $n=0$ or 1 are possible as such powers correspond, respectively, to suppressions by $m_P$ or $m_P/\sqrt{|\xi|}$, both larger than $\Lambda_{UV}$.

(2) While (\ref{S2Vt}) shows that gravitational corrections vanish for $\xi=1/6$ at ${\cal O}(\epsilon_0^2)$ it can be shown that a nonzero contribution appears in $S$ at the ${\cal O}(\epsilon_0^3)$ order $\kappa \beta_\lambda'{}^2$.

(3) It can be checked that $dS/d\phi_0=0$ gives back the constraint (\ref{Const2}), as expected on general grounds.

Numerical ${\cal O}(\epsilon_0^2)$ results for $\phi_0/m_P$ from (\ref{Const2}) and for the tunneling action from (\ref{S2Vt}) are shown with a solid violet line in Fig.~\ref{fig:phi0rungrav} and Fig.~\ref{fig:Sxi}, respectively,  showing good agreement with the fully numerical points. At $|\xi|\sim 20$, the error in $\phi_0/m_P$
is $14\%$ and $1.2\%$ in $S$, a marginal improvement with respect to the ${\cal O}(\epsilon_0)$ approximation of section~\ref{sec:Vrunlgrav}.

It is straightforward to rewrite the previous results for the constraint and action in terms of $\beta_\lambda$ and $\beta_\lambda'$ evaluated at the scale $\phi_0/a$, as we did in Section~\ref{subsec:SMpa}. At this order however, $a=1$ is a better choice and is the one we take. Writing
\be
\lambda_a=-\lambda(\phi_0/a)\ ,\quad
\beta_{\lambda,a}=\beta_\lambda(\phi_0/a)\ ,\quad
\tilde\beta_{\lambda,a}\equiv \beta_{\lambda,a} +\beta_{\lambda}' L_a
\ ,
\ee
where $L_a=\ln a$, the constraint for $\phi_0$ takes the form
\bea
0&=&\tilde\beta_{\lambda,a} -\frac56 \beta_\lambda'+
\frac{12}{5}\lambda_a y_\xi^2\kappa \phi_0^2+\frac{12}{35}\lambda_a\kappa^2\phi_0^4y_\xi^3\left[\frac{809}{25}y_\xi-1\right]\nonumber\\
&+&\frac{1}{25}\kappa\phi_0^2y_\xi\left\{\frac{97}{8}\beta_{\lambda}'-\frac{15}{2}\tilde\beta_{\lambda,a}+y_\xi\left[(101+60L_a)\tilde\beta_{\lambda,a}+\left(10\pi^2-\frac{5777}{60}-30L_a^2\right)\beta_{\lambda}'\right]\right\}
\nonumber\\
&+&
\frac{1}{12\lambda_a}\left[8(3L_a-1)\tilde\beta_{\lambda,a}^2+(19-20L_a-12L_a^2)\tilde\beta_{\lambda,a}\beta_{\lambda}'+(-17+\pi^2+10L_a^2)\beta_{\lambda}'{}^2\right]\ ,
\label{Cons2a}
\eea
where $y_\xi\equiv \xi-1/6$. 

Using the same quantities, the tunneling action reads
\bea
S&=&
\frac{8 \pi^{2}}{3 \lambda_a}+\frac{8 \pi^{2} }{3 \lambda_a^{2}}\left[\frac12\left(\frac{19}{18}-L_a^2\right)\beta_\lambda'+\left(L_a-\frac56\right) \beta_{\lambda,a}\right]+\frac{16 \pi^{2} }{5 \lambda_a}  y_\xi^{2} \kappa\phi_0^{2}\nonumber\\
&+& \frac{\pi^{2}}{9 \lambda_a^{3}}\left\{\left[18 K_0+\frac{553}{36}-\frac{11\pi^{2}}{6}+2\left(\frac{25}{6}-\frac{19}{3}L_a+3L_a^2\right)\right]\beta_\lambda'^{2}\right.\nonumber\\
&+&\left.\left[-\frac{97}{3}+2\pi^{2}+12L_a\left(\frac49+\frac53 L_a-2L_a^2\right)\right] \beta_\lambda'\tilde\beta_{\lambda,a}+\left[\frac{46}{3}+4L_a(-7+6L_a)\right]\tilde\beta_{\lambda,a}^{2}\right\} \nonumber\\
&+& \frac{\pi^{2}}{5 \lambda_a^{2}}y_\xi\kappa\phi_0^{2}\left\{\left[-\frac{419}{90}+\left(\frac{8 \pi^{2}}{3}-\frac{623}{25}-\frac{80}{3}L_a-8L_a^2\right)
y_\xi\right]\beta_\lambda'\right.\nonumber\\
&+&
\left.
\left[\frac{14}{3}+4\left(\frac{1}{15}+12L_a\right)y_\xi\right]\tilde\beta_{\lambda,a}
\right\}
+ \frac{8\pi^{2}}{35 \lambda_a}y_\xi^{3}\kappa^2 \phi_0^{4} \left[6+\frac15\left(\frac{739}{5}+168L_a\right)y_\xi \right] .
\label{S2a}
\eea
The values of $\phi_0/m_P$ from (\ref{Cons2a}) and of the tunneling action from (\ref{S2a}) are plotted respectively in Fig.~\ref{fig:phi0rungrav} and Fig.~\ref{fig:Sxi} (solid blue lines). 
The agreement with the fully numerical results is quite good: for $|\xi|\sim 20$ the error in $\phi_0/m_P$ is $3.8\%$ and $0.14\%$ for $S$, although these numbers represent a marginal improvement over those obtained at ${\cal O}(\epsilon_0)$ in section~\ref{sec:Vrunlgrav}. This shows that, if needed, a better analytical precision should come from a more faithful description of the shape of the SM potential rather than from corrections of higher orders in $\epsilon_0$.

\section{Summary and Conclusions\label{sec:conc}}

This paper revisits the decay of the metastable Standard Model vacuum using an analytic perturbative approach with expansion parameters that correspond to two scale-breaking effects: the running of the Higgs quartic coupling and the presence of gravity (with a nonminimal coupling $\xi$). Analytical results are helpful to grasp the parametric dependence of different quantities (like the tunneling action or the field value of the bounce) and to properly understand numerical results.
The discussion clarifies the conditions under which a proper Euclidean bounce $\phi_B(r)$ describing the decay exists, making use of a very useful integral constraint that $\phi_B(r)$ must satisfy. This constraint, used long ago by Affleck \cite{Affleck} in discussing constrained instantons, follows simply from stationarity of the action and dimensional analysis and proves quite useful to derive analytical expressions for the scale of the bounce in field space, $\phi_0\equiv\phi_B(0)$.
We also highlight the general link between Affleck's integral constraint and the extremality of the tunneling action with respect to changes in $\phi_0$, as given in Eq.~(\ref{link}). 
We illustrate in Section~\ref{sec:VmL} how this approach works using a toy model and then apply it to the case of the SM vacuum decay. First (Section \ref{sec:Vrunl}) we analyze the problem without gravity, modeling the running of the Higgs quartic analytically, so that one is able to obtain analytical results both for the Affleck constraint (that gives $\phi_0$) and then for the tunneling action, at first order in the expansion parameter, taken to be $\beta_\lambda'\equiv d^2\lambda/(d\log\mu)^2$, where $\mu$ is the renormalization scale (that is, $\beta_\lambda'$ is the ``running of the running coupling''). The simple result $\phi_0=e^{5/6}\mu_0$, where $\mu_0$ is the scale at which $\beta_\lambda(\mu_0)=0$, makes precise the usual expectation $\phi_0\sim\mu_0$, which dates back to \cite{Arnold}.

Then we add gravitational corrections in Section~\ref{sec:Vrunlgrav}. At first order in the perturbative expansion (in $\beta_\lambda'$ and $\phi_0^2/m_P^2$) Affleck's constraint leads to a very simple relation that shows the interplay between gravity and running for the determination of $\phi_0$, explaining in particular how gravity lowers significantly $\phi_0$ for increasing values of $\xi^2$. This effect is shown to be related to the fact that the scale at which gravitational effects become relevant is $m_P/\xi$ (as discussed before in the context of Higgs inflation \cite{BLT,BE}). The tunneling action is also calculated at first order and compared with fully numerical results. The analytical result,  improved to take into account that $\phi_0\ll \mu_0$ at large 
$|\xi|$, has a precision better than $0.17\%$ for $|\xi|\leq 20$. 

Once the first order perturbative result is in place and fully under control, it is not difficult to obtain the second order corrections. In order to do that we found it convenient (for its simplicity) to use the tunneling potential formulation of \cite{E,Eg} instead of the usual Euclidean approach. Section~\ref{sec:Vt} applies this new method to the SM potential and obtains in a very simple manner the first-order results of Section~\ref{sec:Vrunlgrav}. Then, Section~\ref{sec:high} calculates explicitly Affleck's condition for $\phi_0$ and the tunneling action at second order in perturbations. The second order result is quite close to the first order one, with a marginal reduction of the error of the analytic evaluation of the tunneling action, which is now below $0.14\%$ for $|\xi|\leq 20$.

%%%%%%%%%%%%%%%%%%%%%%%%%%%%%%%%%%%%%%%%
\section*{Acknowledgments\label{sec:ack}} 
%%%%%%%%%%%%%%%%%%%%%%%%%%%%%%%%%%%%%%%%
I thank Pepe Barb\'on, Alberto Salvio, Stephen Stopyra, Alessandro Strumia and Alfredo Urbano for interesting discussions and/or data sharing.
This work has been supported by  the Spanish Ministry MINECO under grants  2016-78022-P and
FPA2014-55613-P and the grant SEV-2016-0597 of the Severo Ochoa excellence program of MINECO .

\end{document}